\documentclass[a4paper,11pt]{article}
\usepackage[T1]{fontenc}
\usepackage[a4paper,left=3cm,right=3cm,top=2.5cm,bottom=2.5cm,includeheadfoot]{geometry}

\usepackage{lineno}
\usepackage{float}
\usepackage{graphicx}
\usepackage{hyperref}
\usepackage{booktabs}
\usepackage[tight]{subfigure}
\usepackage{caption}
\usepackage{array}

\usepackage{afterpage}
\usepackage{graphicx}

\usepackage{amsmath,amssymb,amsfonts}
\usepackage{siunitx}
\usepackage{algorithmic}
\usepackage{textcomp}
\usepackage{xcolor}
\usepackage[table]{xcolor}
\usepackage{tkz-euclide}
\usepackage{framed}

\usepackage{tikz}
\usepackage{adjustbox}
\usepackage{url}
\usepackage{libertine}
\usepackage{pdflscape}
\usepackage{enumitem}

\usetikzlibrary{positioning, arrows, shapes,chains}
\usepackage{color}
\usepackage{transparent}
\graphicspath{img/}
\usepackage{multirow}
\usepackage{rotating}

\usepackage[framemethod=tikz]{mdframed}
\newmdenv[
  innerleftmargin=6pt,
  innerrightmargin=6pt,
  innertopmargin=6pt,
  innerbottommargin=6pt,
  linewidth=1pt
]{myframe}

\usepackage{cleveref}

\Crefformat{figure}{Figure~#1}
\Crefmultiformat{figure}{Figures~#2#1#3}{ und~#2#1#3}{, #2#1#3}{ und~#2#1#3}
\Crefrangeformat{figure}{Figure~#1--#2}

\Crefformat{section}{Chapter~#1}
\Crefmultiformat{section}{Chapters~#2#1#3}{ und~#2#1#3}{, #2#1#3}{ und~#2#1#3}
\Crefrangeformat{section}{Chapters~#1--#2}

\Crefformat{table}{Table~#1}
\Crefmultiformat{table}{Tables~#2#1#3}{ and~#2#1#3}{, #2#1#3}{ und~#2#1#3}
\Crefrangeformat{table}{Table~#1--#2}

\Crefformat{equation}{Forumla~#1}
\Crefmultiformat{equation}{Forumlas~#2#1#3}{ und~#2#1#3}{, #2#1#3}{ und~#2#1#3}
\Crefrangeformat{equation}{Forumlas~#1--#2}

\begin{document}

\begin{titlepage}
    
    \begin{center}
    
    \vspace*{6cm}
    
    \huge{A Practical Guide to Establishing Technical Debt Management} \\
    \normalsize
    
    \vspace*{1cm}
    \Large{A Whitepaper of \\
    \textbf{Marion Wiese}} \\
    \today
    \vspace{1cm}
    
    \end{center}
    \normalsize
    
\end{titlepage}

\newpage
\renewcommand{\contentsname}{Inhalt}
\tableofcontents

\newpage

\section{Introduction}
\label{sec:introduction}

This white paper provides an overview of the topic of ``technical debt'' and presents an approach for managing technical debt in teams. 
The white paper is based on the results of my dissertation (Wiese et al.~\cite{wiese_it_2023, wiese_techdebt_2025, Wiese2026_Establishing, Wiese2026_bestpractices}), which aimed to translate scientific findings into practical guidance. 
To this end, I collaborated with other researchers to support three teams from different companies in adapting and establishing a technical debt management system tailored to their specific needs. 
Research findings were supplemented with details or additional approaches.
Research results that were less practical were discarded.

The result is a guide on establishing technical debt management within a team. 
The guide is intended to provide orientation and not be a rigid framework. 

We distinguish between ``best practices'' and ``nice-to-haves.'' 
``Best practices'' are understood to be all approaches that were adopted by all three teams. 
``Nice-to-haves'' were used by at least one team.
In many places, it is explicitly mentioned that the team should decide together how to design the process. This also applies, of course, to all areas where this was not explicitly mentioned. 

This white paper explicitly does not cover the establishment of technical debt management across the entire company, but provides suggestions for this at the end.

\subsection{How to read the guide}
\label{sec:howto}
The white paper is written in a way that allows practitioners to quickly grasp the essential aspects and gain initial insights into the topic by skimming through it. In addition, all essential instructions for action have been highlighted in the margins.

\Cref{sec:background} provides a rough overview of the basics of technical debt that are not explicitly part of the guide.
If you are already familiar with the basics of technical debt, you can skip this chapter.

\Cref{sec:guideline} is the core of this white paper and provides guidance on introducing a process for managing technical debt. In this chapter, we draw a fundamental distinction between ``best practices,'' i.e., process elements that have been established by all three teams, and optional additions, i.e., process elements that have been used by some of the teams.

In~\Cref{sec:problems}, we list the problems that typically arise when introducing the process and provide tips on how to avoid them or which aspects require special attention.
For a successful introduction, we recommend reading this chapter as well, as it contains many references to typical errors that might otherwise go unnoticed.

Ultimately, in~\Cref{sec:conclusion}, we provide an outlook on how the process can be extended to an entire company, identify existing obstacles, and discuss the potential advantages this approach could bring.
This chapter is particularly interesting for managers or software architects who are considering a company-wide approach.

\subsection{Terminology}
\label{sec:terms}

This white paper uses some terms without further explanation. We have included an overview of these terms here.
As these are common terms in the field of IT, the list serves primarily to clearly define terms and avoid ambiguities. A computer scientist working in an agile environment should be able to read the text without these explanations.

    \begin{sidewaystable*}[htbp]
        \centering5
        \begin{tabular}{p{4.7cm}p{18.5cm}} 
            \toprule
            Term & Explanation  \\ 
            \midrule 
                (The) Team 
                    & The team for which technical debt management is being introduced. Includes: developers, at least one person responsible for architecture, and the product owner.  \\
                User 
                    & Users of the IT system. In a company, this is usually not the person who pays for the IT system.  \\
                Customer 
                    & Company or an individual who pays for the system.  \\
                Functional  Requirement 
                    & The desire (usually from customers/users) to change system functionalities or add new ones.   \\
                Quality Attribute 
                    & Implicit requirements for the system that are not described by the customer/user but are expected (e.g., performance, availability). The typical quality attributes are described by ISO 25010.    \\
                Non-func\-tion. Requirement 
                    & Requirements for quality attributes and other technical requirements (e.g., use of certain technologies based on company guidelines or existing expertise). \\
                Issue  
                    & Requirement for a system that is stored in a tool. Various attributes are determined or recorded for each issue, e.g., assignee, open date, title, description, effort estimate, and keywords. \\
                Issue Tracking Tool 
                    & A tool that stores issues with their attributes and allows them to be recreated/edited. The issues are displayed as a list or sorted by status in a ``board.'' The list can be sorted and filtered. Typical examples are: Jira, AzureDevOps, GitLab Issues, GitHub Issues. \\
                Issue Types 
                    & There are different types of issues. In addition to technical requirements (e.g., defined as user stories), there are usually bugs/errors. There are also issue types at other hierarchical levels: epics are a collection of issues that are related in terms of content, while tasks or subtasks are components of issues.
                \\
                User Story 
                    & A way of describing functional requirements. Includes the requester, the requirement, and the rationale for the requirement. The term is often used (somewhat inaccurately) in place of the term ``functional requirement.'' \\
                Sprint 
                    & Periods of a few weeks (often 2 weeks). A goal is set for this period and the issues to be dealt with are planned.\\
                Backlog 
                    & Planning tool for collecting issues per system/project. Issues can have different statuses (e.g., new, refined, in progress, in testing, completed).  \\
                Refinement 
                    & Meeting of the entire team to discuss and evaluate new issues. For each issue, implementation ideas are documented, the effort required is evaluated, and other attributes are estimated/determined.   \\
                Planning 
                    & Meeting of the entire team to discuss which issues should be addressed in the upcoming sprint.   \\
                Product Owner (PO)
                    & The person in the team who collects technical requirements, prioritizes them, and has them evaluated by the team during refinement. The product owner determines the order of issues in the backlog and thus their priority.   \\
                DoR -- Definition of Ready 
                    & Defines the requirements an issue must meet to be included in a sprint. For example, estimation is available, acceptance criteria are defined, and dependencies on other teams are clarified.  \\
                SP -- Story Points 
                    & Type of effort estimation; often used in agile environments; an abstract or relative type of estimation. The complexity of the task is compared with the complexity of a similar task and assessed as equal, greater, or lesser. \\  
                PD -- Person days 
                    &  Type of effort estimation; often used in traditional methods; a concrete type of estimation. The effort is estimated in a real unit, i.e., in days. \\  
                Risk (vs. Drawback) 
                    & A risk is a drawback that has a probability associated with it. Example: ``The license costs are very high'' is a drawback. ``We are dependent on the tool, and the license costs could increase'' is a risk.  \\
                CoP -- Community of Practice
                    &  Regular, voluntary meetings of people from a company to discuss a specific topic.  \\
            \bottomrule
        \end{tabular}
        \caption{Terms used in this white paper without further explanation.}
            \label{tab:terms}
    \end{sidewaystable*}

\newpage
\section{Fundamentals of Technical Tebt}
\label{sec:background}
    In this chapter, we provide a brief overview of the relevant basics on the topic of technical debt.

\subsection{Defining the Term Technical Debt}
\label{sec:definition}

    The term was originally established by Ward Cunningham. In his paper, Cunningham wrote: 
    \begin{myframe}
    \textit{``Shipping first time code is like going into debt. A little debt speeds development so long as it is paid back promptly with a rewrite. \ldots The danger occurs when the debt is not repaid.~\cite{Cunningham1992}}
    \end{myframe}

    The term ``technical debt'' is therefore a (technical) metaphor for financial debt, and its terminology is based on terms familiar from the banking sector, such as interest or repayment.
    Interest is the additional expense that arises when technical debt is not repaid. Repayment is the expense that must be incurred to remove technical debt from a system.
    Interest in particular is often forgotten in the discussion about technical debt, or is difficult to grasp.
    
    After some discussion about the exact definition, a group of researchers came up with the following definition:
    \begin{myframe}
    \textit{
    In software-intensive systems, technical debt is a collection of design or implementation constructs that are expedient in the short term, but set up a technical context that can make future changes more costly or impossible. Technical debt presents an actual or contingent liability whose impact is limited to internal system qualities, primarily maintainability and evolvability.~\cite{Avgeriou2016a}}
    \end{myframe}
        
    The part entitled ``future changes'' is particularly relevant here, as it explains two aspects of technical debt:
    \begin{itemize}
        \item Technical debt is only a problem if you plan to change the system or the component affected by the debt.
        If the code is no longer to be changed, for example, in a legacy system that is to be replaced and will no longer be modified, then the technical debt in this code/system does not pose a problem.
        \item Over the past decade, agility has become a major topic in software development. A core element of agility is ``embrace change,'' i.e., welcoming constant change in an IT system. This is also reflected in ``continuous integration'' and ``continuous deployment'' efforts. The code base and everything around it are now changing, in some cases, by the minute. That's why technical debt has become a problem in recent years. The consequences of technical debt become apparent when you want to change a system, which we now do every minute.
    \end{itemize}
    
         \begin{figure*}[!tp]
            \centering
            \includegraphics[width=0.4\textwidth]{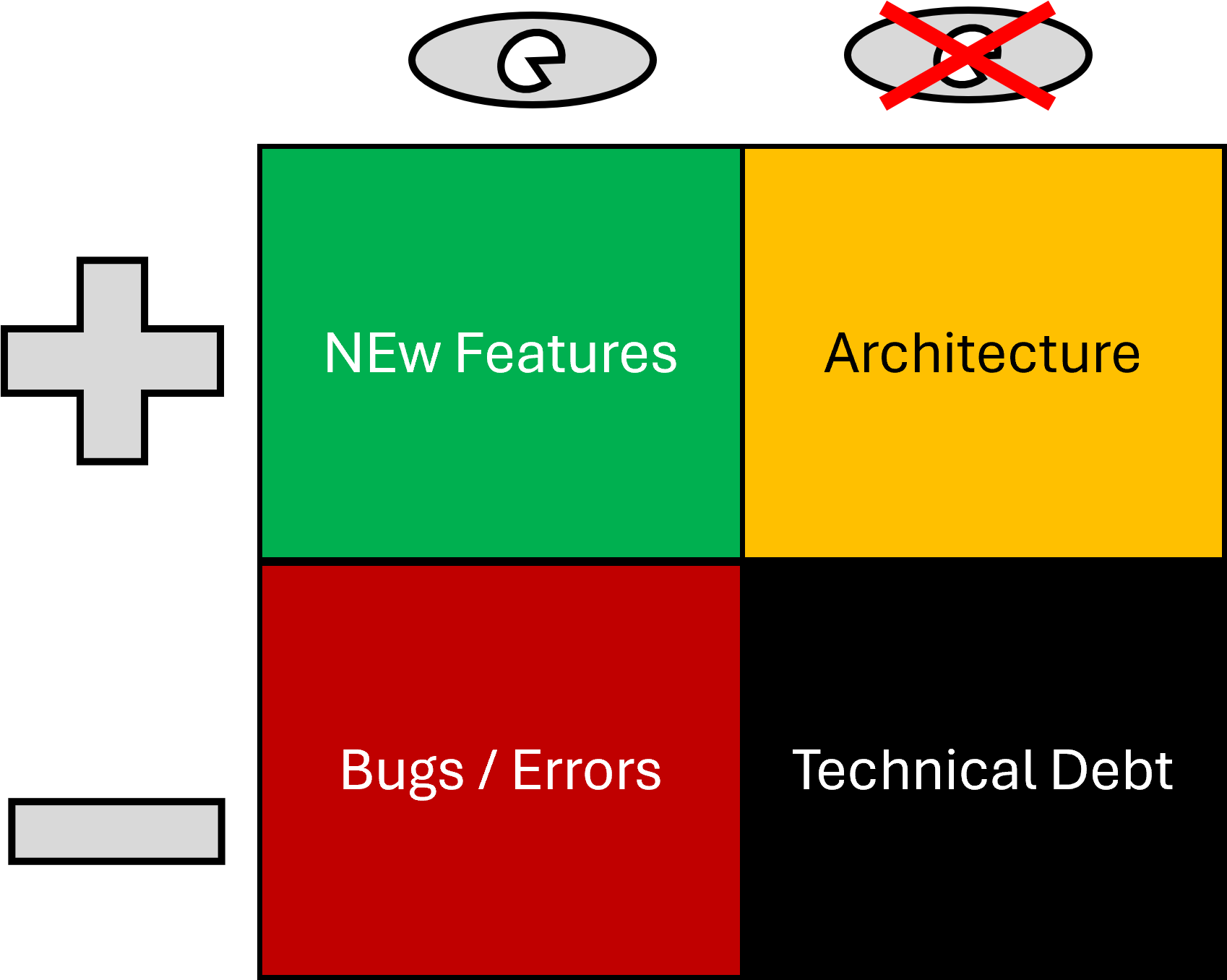}
            \caption{The four colors (classes) of the backlog according to Kruchten et al.~\cite{Kruchten2019}}
            \label{fig:BacklogColors}
        \end{figure*}
        
    Another way to classify technical debt is based on the ``four colors of the backlog'' according to Kruchten et al.~\cite{Kruchten2019}. According to this classification, issues in the backlog of an issue tracking system can be differentiated according to whether they represent a positive or negative value for the system and whether they are visible or invisible to customers or users.
    This results in the classification shown in~\Cref{fig:BacklogColors}.

   In the scientific context, the term ``technical debt item'' has become established as a singular term for technical debt in order to distinguish between a specific problem in the code and a diffuse ``technical debt in the system.''

\subsection{Technical Debt Types}
\label{sec:types}

    Technical debt is often only known as problematic code locations. TD exists, but not only in code, but also in many other software artifacts. \Cref{tab:Types} provides an overview of these different types.

   	\begin{table*} [!htb]
    	    \centering
        	\begin{tabular}{p{3cm}p{10cm}}
    			\toprule
        	   	Type  & Example   \\ 
    			\midrule 
    			Code 	&  Unreadable code; code duplication \\ 
    			Architecture	&  Outdated architectural patterns, such as shared databases\\ 
    			Documentation	&  missing or outdated documentation\\ 
    			Test 	&  Lack of test environment; insufficient test coverage\\ 
    			Infrastructure	& Outdated servers; operating systems not up to date   \\ 
    			Requirements	&  Non-functional requirements not taken into account \\ 
    			Build / Automation	&  No automated deployment; unstructured deployment pipeline \\ 
    			Security	& Passwords in plain text; unsecured interfaces \\ 
    			Social	& Communication barriers caused by the company's organizational structure; error-intolerant corporate culture \\ 
    			Versioning	& Illegible code due to A/B testing or feature flags; problems with or lack of API versioning\\ 
    			Update	& Outdated 3\textsuperscript{rd} party libraries\\ 
    			Hardware	& Use of outdated, less powerful components, e.g., CPUs in the ``embedded'' sector \\ 
                and more & In science, further types of debt are still being identified and named\\
     			\bottomrule
    		\end{tabular}
    		\caption{Various TD types according to Li et al.~\cite{Li2015}}
                \label{tab:Types}
    	\end{table*}

\subsection{Cause-and-Consequences Model of Technical Debt}
\label{sec:causesconsequences}

    Technical debt follows a cause-and-consequences model, i.e., there are factors that can trigger technical debt and consequences of technical debt that can lead to problems and visible symptoms.
    A special feature is that even the absence of action can lead to TD, e.g., if third-party libraries are not updated.

\subsubsection{Visibility}
\label{sec:visibility}

    A fundamental problem with technical debt is that different stakeholders have different perspectives on the situation. \Cref{fig:Visibility} illustrates these perspectives:
        \begin{figure*}[!htb]
            \centering
            \includegraphics[width=0.7\textwidth]{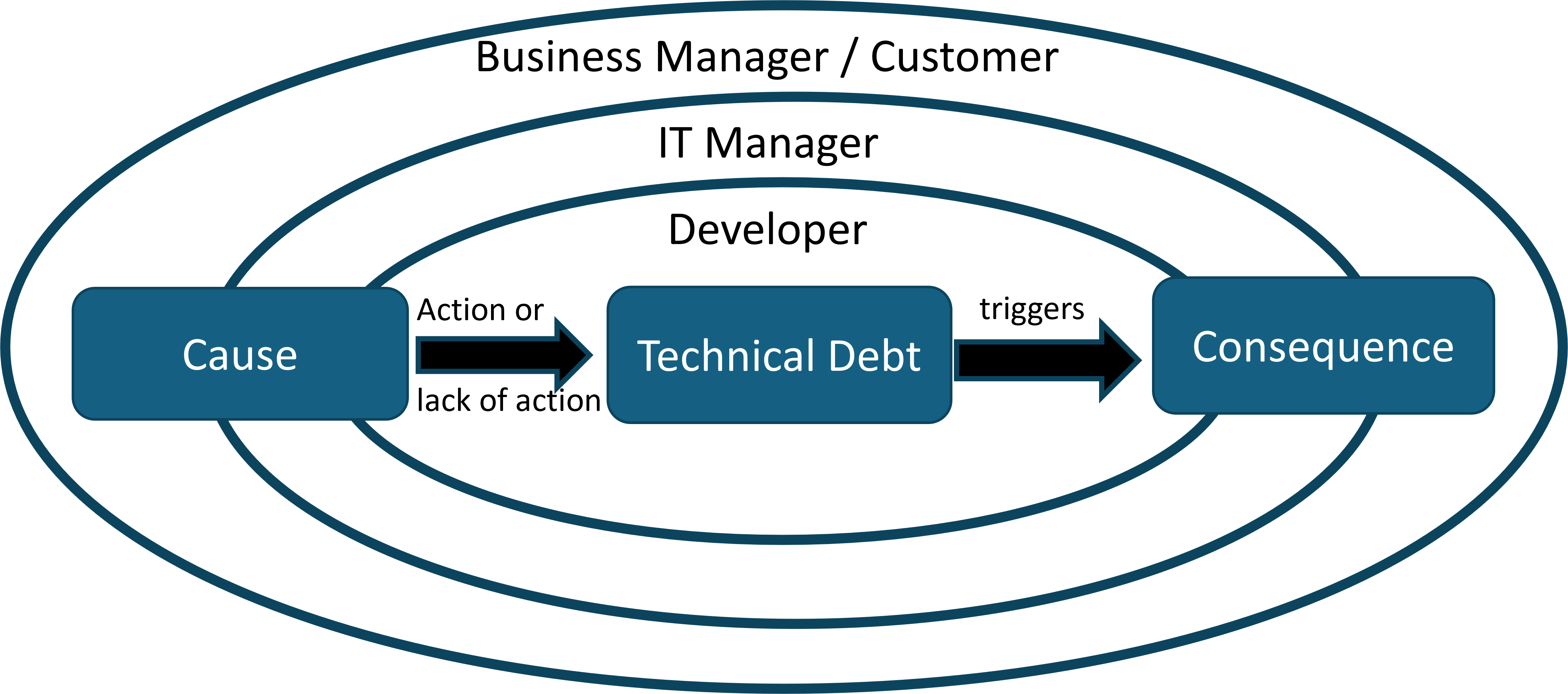}
            \caption{Visibility of technical debt for various stakeholders according to Wiese and Borowa~\cite{wiese_it_2023}}
            \label{fig:Visibility}
        \end{figure*}
        
    \begin{itemize}
        \item \textbf{Developers} have a strong focus on technical debt. They recognize the direct causes, such as deadline pressure or the lack of a specialist with the necessary knowledge. They recognize the direct consequences, such as development taking longer than estimated and missing deadlines. They often lack the bigger picture, i.e., the answers to the questions: Where does the deadline pressure come from? What consequences does the project delay have on the business?
        \item \textbf{IT managers} play a central mediating role. IT managers often understand technical debt themselves, but do not feel its effects directly. They are often much better able to answer the above questions than developers and therefore have a more complete picture.
        \item The \textbf{business managers}, such as customers or technical management, are very familiar with the causes and consequences. They see that they create time pressure. They realize that there are few good developers on the job market. Business stakeholders also see the consequences, e.g., missed deadlines or faulty systems and system failures. The problem is that they do not see the technical debt in the middle and therefore cannot connect the causes and consequences. For them, the middle circle in~\Cref{fig:Visibility} is not visible.
    \end{itemize}

\subsubsection{Categories, Chains, and Cycles}
\label{sec:categorieschainscycles}

        \begin{figure*}[!htb]
            \centering
            \includegraphics[width=0.9\textwidth]{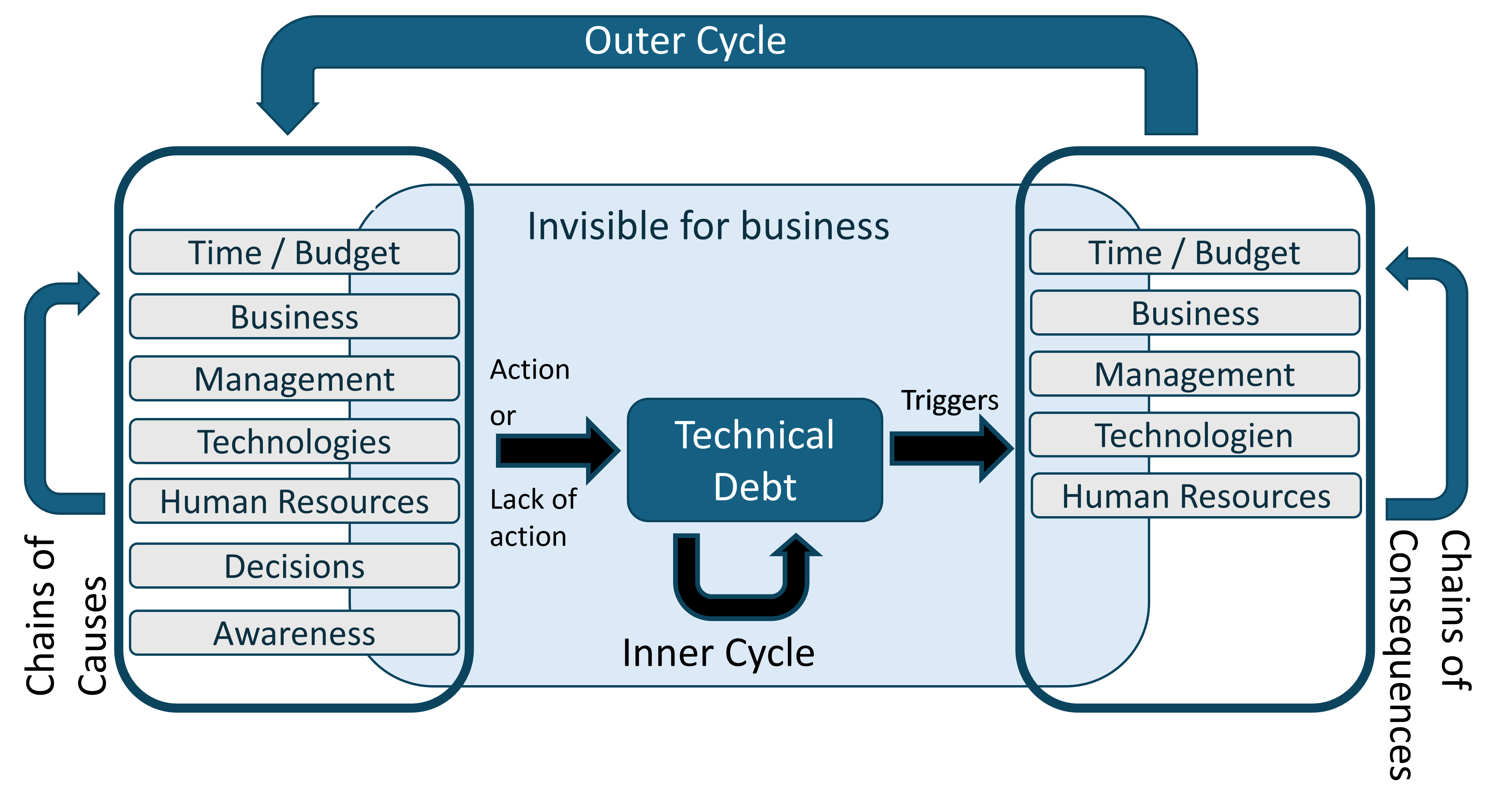}
            \caption{Causes, effects, chains, and vicious circles of technical debt according to Wiese and Borowa~\cite{wiese_it_2023}}
            \label{fig:V4CTD}
        \end{figure*}
                
    The model from~\Cref{fig:V4CTD} shows a more detailed version of the cause-and-effect model. The causes and effects are classified into categories (see~\Cref{tab:categories}). In addition, the arrows indicate that there is usually not just one cause or one effect, but rather that these are often interconnected. Cyclical dependencies are also common. The inner cycles represent the ``broken window'' phenomenon\footnote{\url{https://de.wikipedia.org/wiki/Broken-Windows-Theorie}}, i.e., in an already dirty environment (with broken windows), the probability is higher that further uncleanliness will be added (further windows will be broken). In terms of technical debt, this means that places that already contain a lot of technical debt are most likely to accumulate further technical debt. 
    The outer cycles result directly from the fact that the same categories can be found in both the causes and the effects.
    
   	\begin{table}  [!htbp]
    	    \centering
    	    \footnotesize
        	\begin{tabular}{p{2.9cm}p{5.5cm}p{5.5cm}}
    			\toprule
        	   	Category  & Example of a cause & Example of an effect   \\ 
    			\midrule 
    			Time/Budget 	    & A project deadline that is set too tight & A project cannot meet the deadline  \\ 
    			Business            & The business strategy changes in the middle of the project, affecting the project & Customers are frustrated with product quality and are switching to a competing product \\ 
    			Management	        & Project deadlines are not planned with sufficient buffer time &  Projects are delayed because developers have miscalculated or technical debt must first be cleared\\ 
    			Technology	        & The use of outdated technologies prevents the use of modern methods & Deployment dependencies prevent early rollout of a function that has already been implemented\\ 
    			Human Resources 	& Too many junior developers or developers who are not familiar with the system & Developers are frustrated by the consequences of technical debt and are leaving the company \\ 
    			(Faulty) Decisions	& Management decision to implement a workaround instead of the correct solution; developers do not adhere to common code patterns &  \\ 
    			Awareness	        & If a stakeholder is completely unaware that such a thing as technical debt exists, there is no reason for them to implement or have implemented a more complex solution &  \\ 
     			\bottomrule
    		\end{tabular}
    		\caption{Different categories of causes and effects of technical debt according to Wiese and Borowa~\cite{wiese_it_2023}}
                \label{tab:categories}
    	\end{table}

    Examples:
    \begin{itemize}  
        \item \textbf{Chain of causes:} A new law must be taken into account in the system and has a fixed date on which it will take effect (business -- time). The project manager decides to implement a change in the code in a fixed manner rather than flexibly (management -- decision).
        \item \textbf{Chain of consequences:} During development, a developer discovers that technical debt requires a redesign of a piece of code, changing the estimated effort from 2 days to 2 weeks (technology -- time). The project manager is unable to compensate for this additional effort in the project plan and delivers the desired change with a week's delay (time -- management). The customer is frustrated and considers switching products (time -- business).
        \item \textbf{Inner cycle:} There is no test concept, and no test suite has been selected and is in use. Each new function is not tested, and each new function creates new test debt.
        \item \textbf{External cycle:} A shortage of good developers leads to avoidable problems in the code and reduced readability. Due to high staff turnover, there are only a few developers who are truly familiar with the code. The poor overall situation in terms of implementation quality leads to frequent failures, resulting in overtime and weekend work for developers. Well-qualified developers look for other employers.
    \end{itemize}

\subsubsection{Fowler Quadrant}
\label{sec:fowler}
    M. Fowler provides another simple explanation of the causes of technical debt in his blog\footnote{\url{https://martinfowler.com/}}. The quadrant he has created differentiates between technical debt that is consciously accepted and technical debt that is unconsciously incurred. It also differentiates between careless and wise decisions that can lead to technical debt.
        \begin{figure*}[!htb]
            \centering
            \includegraphics[width=0.8\textwidth]{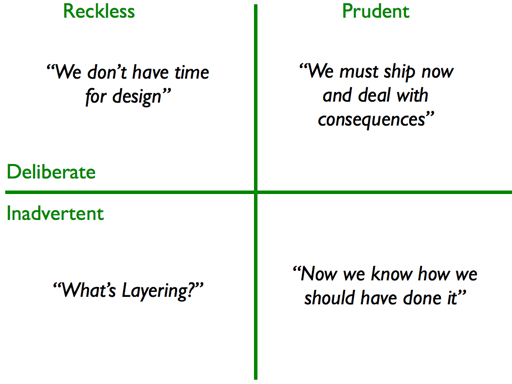}
            \caption{Causes of technical debt according to M. Fowler~\cite{Fowler2009}}
            \label{fig:Fowler}
        \end{figure*}
        
\subsection{Activities for Managing Technical Debt}
\label{sec:activities}
    There are various aspects to managing technical debt. These are briefly explained in~\Cref{tab:activities} and serve as the basis for the guidelines in~\Cref{sec:guideline}.
    
   	\begin{table*} [!htb]
    	    \centering
        	\begin{tabular}{p{3cm}p{10cm}}
    			\toprule
        	   	Activity          & Explanation   \\ 
    			\midrule 
    			Identification 	   &  Identifying technical debt, e.g., through static code analysis, but also determining whether an issue from the backlog is a technical debt \\ 
    			Prevention 	       &  Prevent technical debt from being incurred, e.g., by making more conscious decisions and preventing carelessness \\ 
    			Documentation 	   &   Collecting information on each technical debt (e.g., description, list of tasks for dismantling, assessment of consequences) in order to gain an overview and be able to evaluate the technical debts \\ 
    			Measurement 	       &  Determining or estimating quantifiable characteristics of technical debt (e.g., dismantling costs or interest burden) in order to be able to evaluate the technical debt  \\ 
    			Prioritization 	   & Use the assessment (measurements) of technical debt to prioritize it against each other or against functional requirements  \\ 
    			Repayment 	   & Methods for structuring and sensibly organizing the repayment of technical debt  \\ 
    			Monitoring (\&Visualization) &  Monitoring the system status in terms of technical debt using the recorded and measured properties \\ 
     			\bottomrule
    		\end{tabular}
    		\caption{Activities for managing technical debt according to Li et al.~\cite{Li2015}}
                \label{tab:activities}
    	\end{table*}

\newpage


\section{Guide to Technical Debt Management}
\label{sec:guideline}

    In this chapter, we first explain a few organizational prerequisites for managing technical debt (\Cref{sec:orga}). We then describe how to avoid technical debt (\Cref{sec:prevent}) and the elements needed to manage it (\Cref{sec:manage}). Finally, we provide step-by-step instructions on how to get started with the process (\Cref{sec:gettingstarted}).

\subsection{Organizational Matters}
\label{sec:orga}
    Two key considerations for the organization are: firstly, the personnel requirements (\Cref{sec:TDManager}) and secondly, the technological requirements (\Cref{sec:tool}) for the process.

    \subsubsection{Responsibilities -- TD Manager}
    \label{sec:TDManager}

    \begin{myframe}  
        From the outset, there should be a person responsible for the process who acts as a so-called ``technical debt manager'' (``TD Manager'').
    \end{myframe}
    
    The tasks of the TD Manager are:
    \begin{itemize}  [itemsep=0pt]
        \item Coordination of the implementation process
        \item Technological responsibility for making changes to the issue tracking tool (changing existing issue types, creating new issue types)
        \item Technological responsibility for coordinating visualization and interfaces
        \item ``Voice of reason,'' who repeatedly reminds everyone of the process and also considers the issue of ``technical debt'' in all discussions (e.g., to avoid it in decision-making situations). 
        \item Contact person for technical debt for other teams within the company and participation in a community of practice for this topic.
    \end{itemize}

    \subsubsection{Tool Support}
    \label{sec:tool}
    
    \begin{myframe}
        It is advisable to use the tool that is also used for work organization in everyday life. In an agile environment, this is the issue tracking tool used for the backlog (e.g., Jira, Azure DevOps, GitLab Issues).
    \end{myframe}
    
    \begin{myframe}        
        In addition to issue types for \textbf{functional changes} (e.g., user stories) and \textbf{bugs}, the selected issue tracking tool should also include a separate type for \textbf{technical debt}.
    \end{myframe}
    Other issue types are possible, but they are usually at a different level of abstraction, such as epics as a bracket around several issues or subtasks, and test cases as parts of an issue.
    In the following chapters, we describe how the different issue types should be changed to enable technical debt management.
    
    Since issue tracking tools currently do not offer good support for visualization, it may be necessary to use an additional visualization tool (e.g., Microsoft Power BI\footnote{\url{https://www.microsoft.com/de-de/power-platform/products/power-bi?market=de}}, https://www.microsoft.com/en-us/power-platform/products/power-bi/, 
    Tableau\footnote{\url{https://www.tableau.com/de-de}}).https://www.tableau.com/). 
    It is advisable to ask your company which tool(s) are already in use there.

\subsection{Preventing Technical Debt}
\label{sec:prevent}

    Before addressing the management and repayment of technical debt, it is best to try to prevent it as much as possible.

    To prevent technical debt, it is essential to examine its underlying causes more closely.
    As can be seen from the causes of technical debt, it often arises as a result of a decision (see~\Cref{sec:causesconsequences}). However, this decision can be made consciously or unconsciously (see~\Cref{fig:Fowler}). 

    \begin{myframe}
        The goal of avoidance is therefore not only to make better/smarter decisions (\Cref{tab:EntscheidungenKlug}), but also to be aware of the decisions (\Cref{tab:EntscheidungenBewusst}).
        These decisions are not made when discussing and evaluating technical debt issues, but rather for all issues, especially those issues that aim to make functional changes. 
        To avoid technical debt, it therefore makes sense to adapt the issue template for functional requirements.
    \end{myframe}
    
    The bug template or epics can also benefit from this change, but this depends heavily on how these issue types are used.
    
    Below, we present adjustments that help the team avoid technical debt \textbf{in all issue types}. 
    
    For factors that serve to make the decision conscious (\Cref{tab:EntscheidungenBewusst}), checkboxes are sufficient, as the aim is simply to remind the team of certain facts. Instead of a checkbox, the question can also be included in a Definition of Ready (DoR) if this is used consistently by the team.
    
    For the factors in decision-making (\Cref{tab:EntscheidungenKlug}), it is up to the team to decide whether the findings are stored as free text in the issue or whether only a check box is used to remind them to discuss these topics.

       	\begin{table*} [!htb]
    	    \centering
    	    \footnotesize
        	\begin{tabular}{p{2.3cm}|p{9.5cm}|p{1.8cm}}
    			\toprule
        	   	Attribute  & Explanation  & Field / Attribute type \\ 
    			\midrule 
        	   	Talked about TD   
                    & The check box serves as a reminder that technical debt should generally be discussed in the context of the issue. It is checked off when the team determines that the issue does not generate technical debt, or after an issue of the type ``technical debt for dismantling'' has been created for deliberately incurred technical debt.  
                    & check box \\ 
    			\midrule 
        	   	Is this issue a repayment of technical debt? 
                    & This question serves as a safeguard for teams that cannot quickly and easily determine whether new issues involve technical debt reduction. If the question is answered with ``No,'' it is marked as answered. If the question is answered with ``Yes,'' the type and template for this issue are switched.  
                    & DoR or check box  \\ 
        	   	Does this issue create technical debt? 
                    & This check box serves as a reminder that the question should be answered during the discussion of the issue. If the question is answered with ``No,'' the question is marked as answered. If the question is answered with \textbf{``Yes'}, a discussion will take place as to whether it is wise to accept technical debt here (see~\Cref{tab:EntscheidungenKlug}) and, if necessary, \textbf{another issue for dismantling this technical debt} will be created, evaluated, and linked to this issue. 
                    & DoR or check box \\ 
     			\bottomrule
    		\end{tabular}
    		\caption{Methods for making conscious decisions regarding technical debt. (Use of the first or the last two attributes)}
                \label{tab:EntscheidungenBewusst}
    	\end{table*}

       	\begin{table*} [!htb]
    	    \centering
    	    \footnotesize
        	\begin{tabular}{p{2.3cm}|p{9.5cm}|p{1.8cm}}
    			\toprule
        	   	Attribute  & Explanation  & Field / Attribute type \\ 
    			\midrule 
    			Quality attributes 	
                    & For each issue, you should consider whether a quality attribute is particularly affected by the implementation. If quality attributes are impaired, e.g., performance could deteriorate, this provides a direct transition to the next point.
                    & label field or free text  \\ 
    			Drawbacks 
                    & Discussing and, if necessary, recording the drawbacks of a solution helps to raise awareness of these drawbacks so that they can be addressed in the workplace. It also helps to think about other solutions (alternatives, see below) that do not have these drawbacks. 
                    & Check box or free text \\ 
    			Risks 	 
                    &  Discussing and, if necessary, recording the risks of a solution is helpful in developing mitigation strategies for these risks. However, it is also helpful to identify other solution options (alternatives, see below) that do not carry these risks. 
                    & Check box or free text \\ 
    			Alternatives
                    & Comparing different alternatives helps you decide on a solution that carries the least risk of creating technical debt, either immediately or over time. 
                    &  Check box or free text \\ 
     			\bottomrule
    		\end{tabular}
    		\caption{Methods for making wise decisions regarding technical debt}
                \label{tab:EntscheidungenKlug}
    	\end{table*}

    \subsubsection{Indicators of Unconscious Incurrence of Technical Debt}
    \label{sec:Unconsciousindicators}

        When refining or planning issues, there are various indicators that may suggest that there is a risk of (unconsciously) incurring technical debt.
        We have included a list of such indicators here, which can help readers remain alert in certain situations and, if necessary, steer the team's discussion toward this issue.

        Statements that indicate unconscious incurrence of technical debt:
        \begin{itemize} [itemsep=0pt]
            \item ``I don't quite understand that yet, but it's not important anyway.''           
            \item ``If we can't do it that way, we'll have to see how it fits.‘’
            \item ``First, let's come up with a small solution.‘’ 
            \item ``We'll do that later.‘’
            \item ``That is extra bells and whistles.'' (i.e., some special solution for a unique situation/customer)
        \end{itemize}

        Circumstances that indicate the unconscious incurring of technical debt:
        \begin{itemize} [itemsep=0pt]
            \item The content of the issue is not yet clear, but the issue will be included in the sprint. Specifications still need to be clarified.
            \item The solution approach for the issue is not yet clear, but the issue will be included in the sprint. 
            \item The process is not being followed; issues that have not (yet) been discussed in refinement are being included in the sprint
            \item The developer who is supposed to implement the issue is not involved in the refinement.
            \item Acceptance criteria are not specified.
            \item No planning poker was performed, resulting in a distorted and possibly too low estimate.
            \item The product owner decides on the effort or provides their own estimate to the developers, resulting in a distorted and possibly too low estimate.
            \item Creation of a ticket for testing so that the functional ticket can be closed even though the test is still missing.
            \item Recognition that technical debt is being incurred (e.g., code duplication), but discussion of this is discontinued.
            \item Decisions in favor of a technology, even though the requirements are not yet clear.
            \item Providing information to users, even though the technical details have not yet been clarified. (Is this possible within the time frame? Is it even possible at all?) 
            \item Installation of special features, wanting to fulfill all user requests, even if they are very complex, not disclosing the financial consequences of their requests to users.
        \end{itemize}

\subsection{Identifying Technical Debt}
\label{sec:identify}

    The first and challenging task in managing technical debt is determining whether an issue is technical debt in the first place. 
    
    The definitions presented in~\Cref{sec:background} can help with this, but they do not apply to all practical situations, for example, when the team does not have direct customers but provides a central service for other development teams. It is also usually not enough to simply look at whether the team incurs any costs (``additional costs'‘, ``interest’') as a result of not implementing the issue.

\begin{figure*}[!ht]
        \centering
        \includegraphics[width=0.3\textwidth]{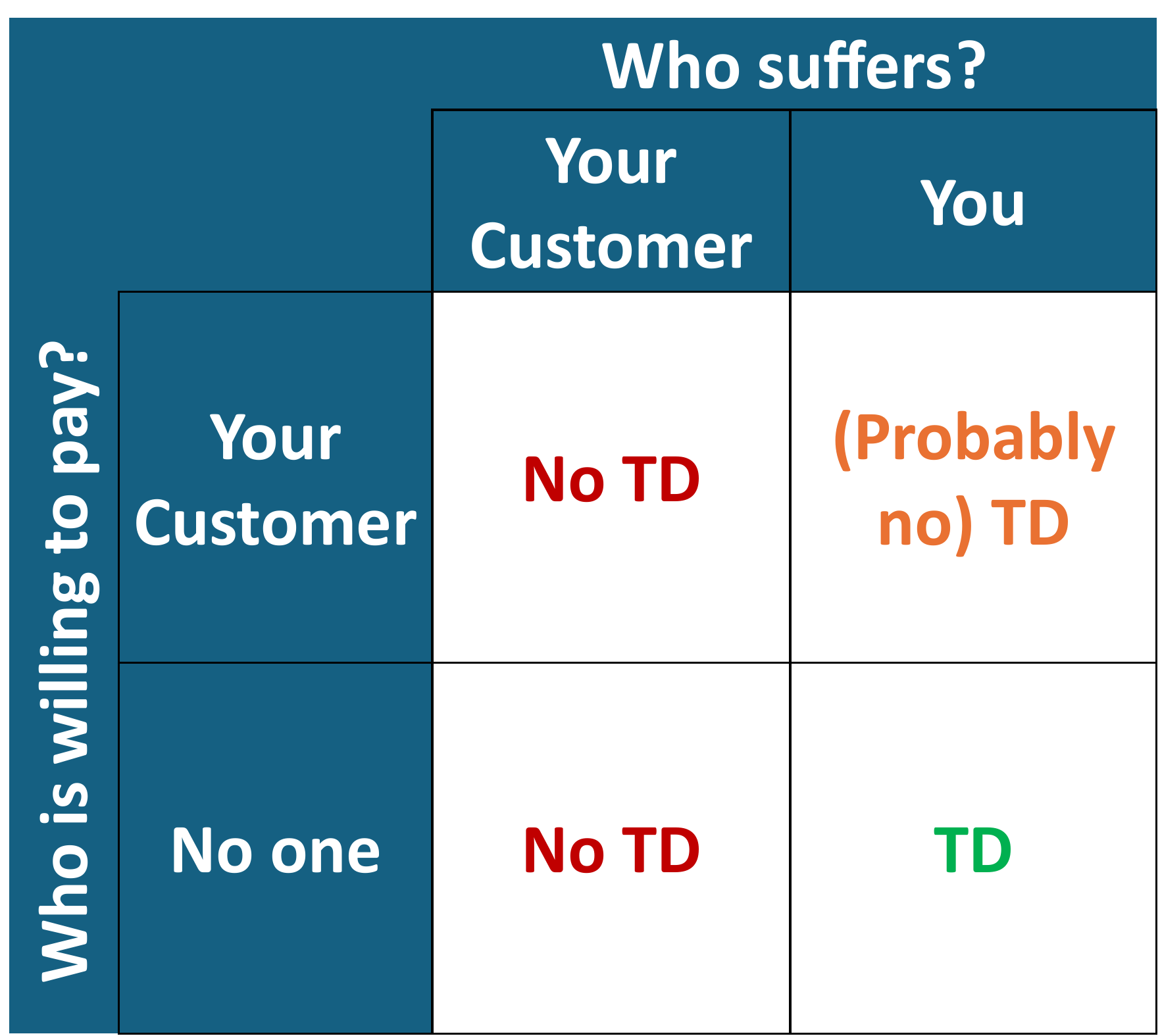}
        \caption{A decision matrix for identifying technical debt}
        \label{fig:matrixsmall}
    \end{figure*}

    \begin{myframe}        
        One method that works relatively well and is easy to use is to focus on two questions:
        \begin{itemize} [itemsep=0pt]
            \item Who wants this issue to be addressed? (Who is suffering from the issue?)
            \item Who is willing to pay for the issue to be addressed?
        \end{itemize}
        \Cref{fig:matrixsmall} shows a matrix with the resulting classification.
    \end{myframe}

\begin{figure*}[!ht]
        \centering
        \includegraphics[width=0.6\textwidth]{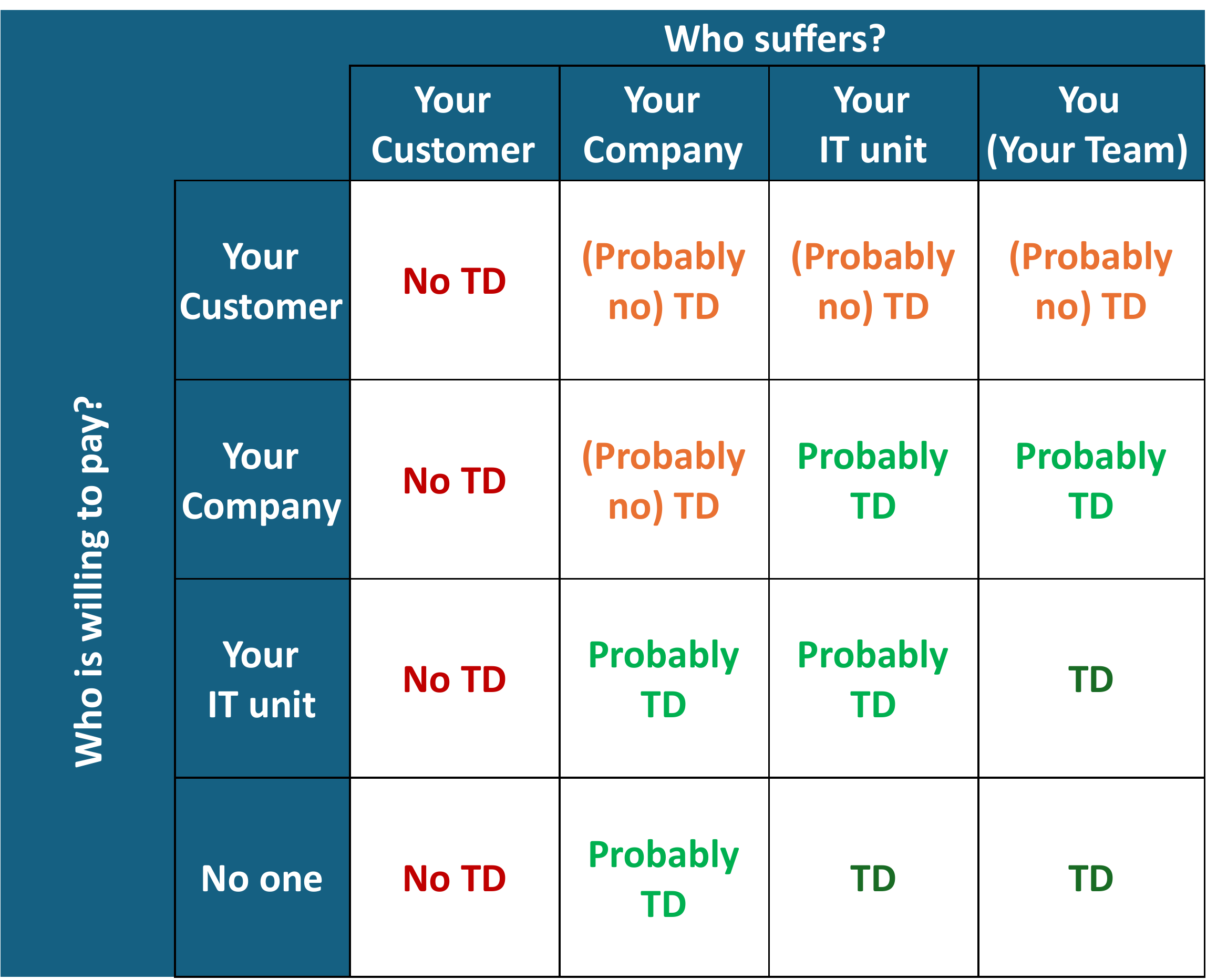}
        \caption{A decision matrix for identifying technical debt, including various stakeholder perspectives}
        \label{fig:matrixbig}
    \end{figure*}

    However, when scaling this matrix to include multiple stakeholders, not all decisions are straightforward, as shown in~\Cref{fig:matrixbig}. 
    The team has to decide what to interpret as TD and what not to. The presented matrix can support this discussion.

\begin{figure*}[!ht]
        \centering
        \includegraphics[width=\textwidth]{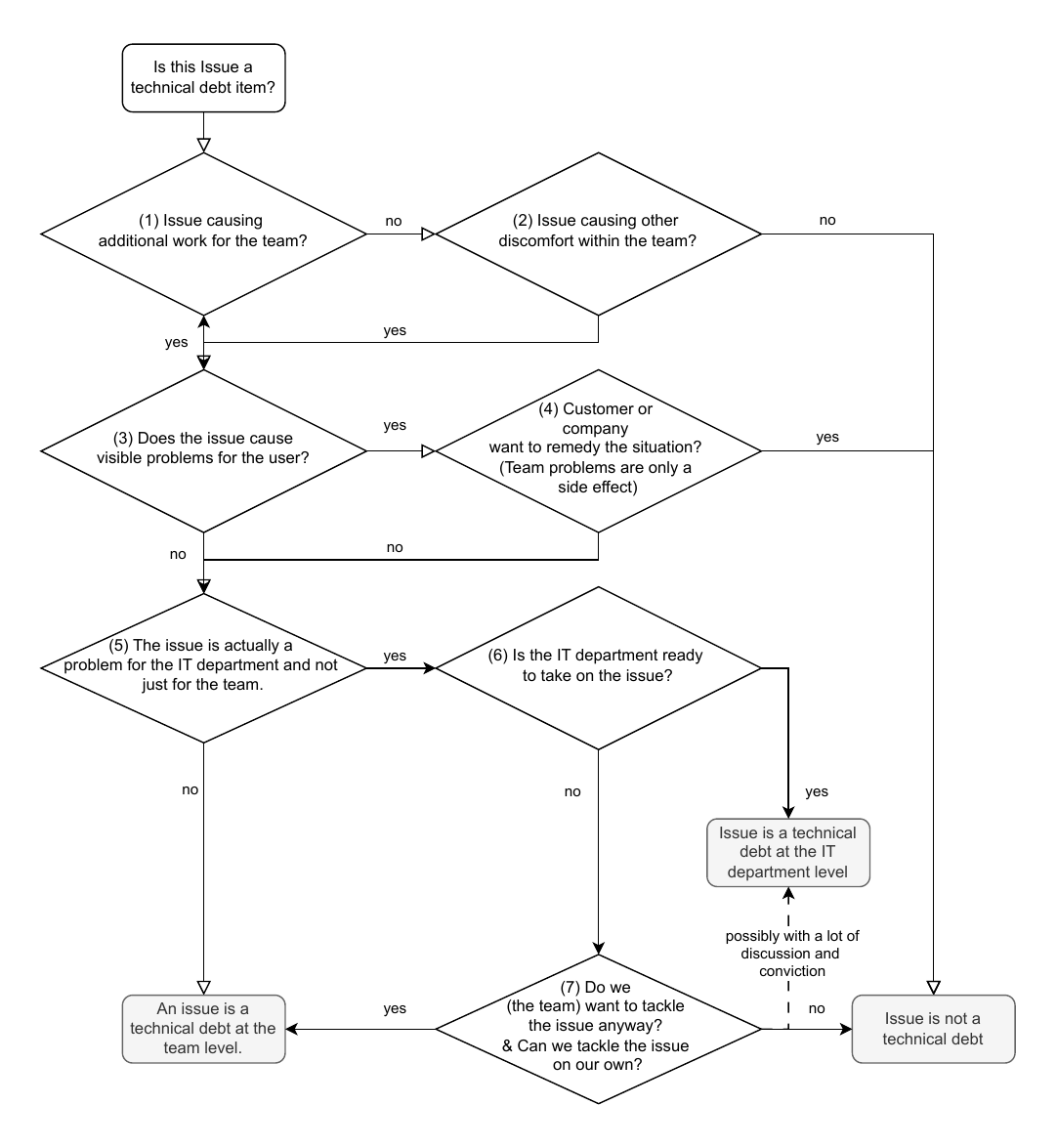}
        \caption{Identifying technical debt (TD), an orientation}
        \label{fig:Identification}
    \end{figure*}

    In addition, the simple flowchart in~\Cref{fig:Identification} helps with classification.
    We begin with the fundamental idea that resources for reducing technical debt are scarce. 
    If you imagine the resources for the different issue types as pots, you want as little as possible to end up in your own pot for technical debt.
    This raises the question of whether -- even if the team would benefit from reducing technical debt --another pot could be used to pay for it.
    This also means that technical debt can be more clearly assigned to different hierarchical levels (at the team level vs. at the IT department level).
    
    To illustrate the branches, example situations for each decision are listed here (see numbering in \Cref{fig:Identification}):
    \begin{enumerate}
        \item Code is difficult to read, which means that further development in this part of the code takes longer. Every morning, a manual production step has to be carried out because there is no automation. Special case: If the additional work needs to be carried out by another IT team (e.g., administrators or support team), this can still be answered with ``Yes'' in order to do the other team a favor, for example.
        \item Error messages repeatedly appear in the log, which do not cause any additional work for the developers. Nevertheless, the developers feel disturbed by this because they would like to have a clean log.
        \item The system's performance is poor when booting. Users simply get their morning coffee during the boot process. However, developers must reboot the system several times an hour if they make changes to the boot process.
        \item The error messages in the log refer to error messages in the interface that the user has also seen. Users may not have reported the error (yet) because they can simply click away the error message. The customer (PO) decides to expand this error despite the lack of complaints from users.
        \item The database used by all IT teams would need to be updated to utilize the new functionalities.
        \item To standardize the tool landscape, the company decides that all teams should switch from their issue tracking tool to Azure DevOps.
        \item The team needs a monitoring tool for secure maintenance. The IT department has no central guidelines for this and no licenses for any tools in stock.  The team decides to use Grafana and purchase the licenses from its own budget.
    \end{enumerate}

    \subsubsection{Static Code Analysis}
    A static code analysis tool, such as SonarCloud or NDepend, can help identify technical debt. However, these tools currently have significant limitations:
    \begin{itemize} [itemsep=0pt]
        \item  This method only finds technical debt in the source code, while other types of debt are not taken into account. 
        \item Many supposed technical debts are found that are not debts at all, so-called ``false positives.'' These false positives cause an effort to identify the actual debts in the mass of reports.
        \item The relevance of the technical debts found is often low, and it is not clear whether they cause any damage at all. The debts identified directly by the developers, on the other hand, have always already caused additional effort for them.
    \end{itemize}
    
    Nevertheless, static code analysis can be a valuable addition to the process described here, as the technical debts identified can be recorded as issues and incorporated into the process.
        
    \subsubsection{Indicators for Technical Debt}
    \label{sec:TDindicators}

    In addition to the matrices presented above, there are various indicators for technical debt. Indicators suggest that technical debt may be present. However, the issues are not necessarily technical debt; the final decision rests with the team and is based on their experience and expertise. Below, we present the indicators in bullet point form:
    
    \begin{itemize} [itemsep=0pt]
        \item The use of certain terms in the title of the issue. The following terms (and their linguistic variations) are indicators:
        \begin{itemize}  [itemsep=0pt]
        \item \textit{Optimizing, improving, revising, testing, documenting, performance, upgrading, updating, refactoring, cleaning up, converting}
        \end{itemize}
        \item Discussing the issue: 
        \begin{itemize}  [itemsep=0pt]
            \item ``We did it quickly back then.'' 
            \item ``We shouldn't have done it that way.'' 
            \item ``Now we want to do it right.'' 
            \item ``We implemented it hard-coded at first.'' 
            \item ``Now we're doing it right.'' 
            \item ``I forgot to expand it back then.''
            \item ``At the time, we thought that would be good.'' 
            \item ``It turned out that this is not optimal.''
            \item ``Special cases would then be eliminated if we implement this.''
            \item ``I don't understand what you're talking about, but it doesn't matter.'' (from the PO)
        \end{itemize}
        \item The issue is not presented by the product owner, but by a developer or architect.
        \item Adapting cross-cutting concerns (e.g., testing procedures, logging mechanisms, monitoring)
        \item Use of new or changes to work tools (e.g., monitoring, documentation, issue tracking, error tracking)
        \item Fixing an error or simplifying a task for which a workaround already exists
        \item Moving a functionality/method to another area (class/package/...)
        \item Improvements to resilience (e.g., adjusting timeout)
        \item Product owner does not understand the issue. 
        \item Communicating the solution and time frame to the user before consulting with the developers
    \end{itemize}

    \subsubsection{If in doubt...}

    If the team is unsure whether an issue constitutes technical debt, it makes sense to reflect on the purpose of the classification.
    We classify an issue as technical debt so that we can then use a different template. This template serves the purpose of making the costs of technical debt visible and prioritizing technical debt. 
    
    \begin{myframe}
        So if it makes sense to evaluate the issue in question (e.g., to justify implementation) or prioritize it (e.g., to rationalize discussions about its importance), then it makes sense to classify it as technical debt.
    \end{myframe}

\subsection{Managing Technical Debt}
\label{sec:manage}

In this chapter, we provide an overview of the components of technical debt management: documenting (\Cref{sec:document}), prioritizing (\Cref{sec:prioritize}), repaying (\Cref{sec:repay}), and visualizing (\Cref{sec:visualize}).

\subsubsection{Document Technical Debt}
\label{sec:document}

    \begin{myframe}
        As with all IT project planning, assessing technical debt can only be based on expert estimates. 
        
        Various attributes have been established for the team to assess when evaluating technical debt.
        We distinguish here between ``best practices,'' i.e., attributes that should always be assessed, and ``nice-to-haves,'' i.e., attributes that can be additionally assessed depending on the team and/or company. 
    \end{myframe}
    
    The~\Crefrange{tab:attributesI}{tab:attributesII} on the following pages list these attributes together with their explanations, field types, and, where applicable, contents, with the ``nice-to-have'' attributes highlighted in gray. 
    
    \begin{sidewaystable*}[htbp]
        \centering
        \begin{tabular}{p{3cm}|p{6.5cm}|p{6.5cm}|p{4cm}}
            \toprule
            Field/attribute &  Explanation  & Example & Field/attribute Type \\ 
            \midrule 
            Risks of non-repayment (benefits of repayment)
                & What can happen if this technical debt is not repaid? There can be several consequences for technical debt at the same time. It is essential to list all consequences (including drawbacks and risks) here. You can use the 5 Whys method for this (see~\Cref{sec:mistakes}).  
                & Additional work for developers, risk of errors in the system, performance losses, and security risks.
                & Free text; content: bullet point list \\ 
            \midrule             
            Repayment effort (SP)	
                &  For planning in the sprint, an effort estimate should be made in the unit that is also used for other issues. 
                & Story points or T-shirt sizes
                & Depending on content: Integer / Text / Combo box  \\ 
            \midrule 
            Contagiousness 	
                & Does the cost of repayment increase over time? 
                & An API with technical debt initially has only one consumer that needs to be changed, but over time, additional consumers are added, all of which then also need to be changed.   
                & Combo box with ``will decrease,'' ``will remain the same,'' and ``will increase'' as options\\ 
            \midrule 
            Interest probability or frequency 	
                & How likely is it that interest will actually be incurred? In principle, this assessment should be made for all ``risks of non-repayment.'' A pragmatic approach that has become established is to assess the most serious risk of non-repayment. 
                & How likely is it that someone will need the missing documentation? (1x/year) How likely is it that the technical debt will result in a production error that causes additional work? (1x/quarter) How often does a workaround have to be performed (manually) by the developers? (daily)
                & Combo box: 1x/day; 1x/week; 1x/month; 1x/quarter; >=~1x/year(half year)  \\ 
            \midrule 
            Interest 	
                & For the most serious risk of non-repayment, an estimate is made here of how much effort will be required in development if the case occurs and interest has to be paid.  
                & (Similar to the above): 1 day of additional effort to understand the code without documentation. 4 hours of effort to find and fix the bug and deploy the hotfix. The workaround requires 15 minutes of a developer's working time. 
                & Combo box: 15~min.; 1~hr.; 4~hrs.; 1~day; >=~2~days\\ 
            \midrule
            Priority 	
                &  see~\Cref{sec:prioritize}
                & similar to other issue types
                & Combo box with 5-point scale \\
            \bottomrule
        \end{tabular}
        \caption{Useful attributes for a ``technical debt'' issue -- Part I (The ``nice-to-have'' attributes are highlighted in light gray)}
            \label{tab:attributesI}
    \end{sidewaystable*}
    
    \begin{sidewaystable*} [htbp]
        \centering
        \begin{tabular}{p{3cm}|p{6.5cm}|p{6.5cm}|p{4cm}}
            \toprule
            Field/attribute &  Explanation  & Example & Field/attribute Type \\ 
            \midrule 
            Resubmission date	
                &  see~\Cref{sec:prioritize}
                &
                & date field \\ 
            
            \midrule 
            Component 	
                &  The (architectural) component (system part, program part) in which the technical debt is located or has an impact.
                & Technical debts in the checkout process of the online shop, in the tax calculation of the booking system, or even in the front-end, during database access.
                & Hierarchy (usually already included in the standard)\\ 
            \midrule  \rowcolor{gray!20}
            Pain Factor 	
                In addition to the hard factors, it sometimes makes sense to consider the developers' subjective assessment. 
                & There is a workaround that is also economically viable, but this recurring task can still be frustrating for developers.
                & Integer or combo box (scale from 1 to 5)\\ 
            \midrule  \rowcolor{gray!20}
            Quality attributes  	
                & List of influenced quality attributes. This information can be helpful if certain quality attributes are particularly important to the business or for planning quality initiatives.
                & In a bank: Technical debt that negatively impacts security. In the embedded domain: Technical debt that negatively impacts performance. 
                & Keyword field, if possible, otherwise free text\\ 
            \midrule  \rowcolor{gray!20}
            Repayment effort (PD)  	
                &  To determine a return on investment (ROI) value (see~\Cref{sec:prioritize}), it is necessary to estimate the effort in person-days (PD) as well.
                & 
                & Numbers / Integers \\  
            \midrule \rowcolor{gray!20}
            Risks of repayment	
                & Technical debt increases the risk of side effects in further developments. Technical debt is often found in system components with a significant amount of other technical debt, resulting in a high risk of side effects. Such risks are listed here; if necessary, their priority can be lowered. 
                & Risk that other parts of the system cannot be further modified during this time; impact on other teams (e.g., when changing an API)
                & Free text; content: bullet point list \\ 
            \midrule  \rowcolor{gray!20}
            Breaking Change  	
                &  As an alternative to the ``Repayment risks'' field, you can simply use a check box to indicate that the issue may have potential side effects. 
                & see above
                & Check box\\ 
            \bottomrule
        \end{tabular}
        \caption{Useful attributes for a ``technical debt'' issue -- Part II (The ``nice-to-have'' attributes are highlighted in light gray)}
            \label{tab:attributesII}
    \end{sidewaystable*}

\subsubsection{Prioritize and Resubmit Technical Debt}
\label{sec:prioritize}

    \paragraph{Determine Priority} 
    Priority is usually measured on a scale of five values, with the names for the levels varying. In some cases, a number between 1 and 5 is assigned. Alternatively, priorities can be labeled as ``very low'' to ``very high.'' Some companies have established more specific labels, such as ``guaranteed,'' ``blocker,'' ``critical,'' ``medium,'' and ``low.'' 
    
    \begin{myframe}
        Prioritization is usually done on a 5-point scale.
        No new scale should be invented for technical debt; instead, the existing scale should be used.
        Since no best practices have been established yet, we present three variants for determining priority: ``educated guess,'' ``mean value,'' or ``return on investment.''
    \end{myframe}
    
\newpage
    \textbf{\textit{Educated Guess:}}
    In this case, the priority is determined jointly by the team. Compared to the previous approach, however, the information required to set the priority in an informed manner is available. 
    
    It is essential that team members agree on the relevant attributes that should be considered when prioritizing.
    For example, the expectation that a situation will occur (e.g., new corporate/IT strategy, new developments in the IT market) could be taken into account. However, it has been found that it makes more sense not to take this expectation into account, but instead to set the resubmission date accordingly (see the section on ``Resubmission'' below). 
    Another example is the consideration of repayment costs. 
    These cannot be taken into account in the ``educated guess'' either if there is a ``low-hanging fruit'' visualization (see~\Cref{fig:VisualisierungPlanung}). With the help of this visualization, the combination of priority and effort can then lead to an order for repaying technical debt items.

    This is the ``simplest'' and most pragmatic option. However, it is also a very subjective option, i.e., subject to the developers' subliminal preferences when making the assessment, and therefore possibly less reliable than the following options.

    \textit{\textbf{Mean value:}}   
    
    In this case, all attribute values are assigned a value between 1 and 5, and the mean value of these values is then used as the priority. An overview of the assignment is provided in~\Cref{tab:meanIntegers}.
    Not all of the listed attributes need to be used to calculate the mean value. The decision on this is up to the team.

    \begin{table*}[htbp]
        \centering
        \begin{tabular}{p{2.5cm}p{3cm}p{2cm}p{5cm}}
            \toprule
            Attribute & Value & assigned Integer & Remark \\ 
            \midrule 
               Interest        & 15 min. & 1 &\\
                             & 1 hr.  &  2 &\\
                             & 4 hrs.. &  3 & \\
                             & 1 day &  4 &\\
                             & $\geq$ 1 day & 5  \\
            \midrule 
               \multirow{2}{2.5cm}{Interest probability }  & $\geq$ 1x/year & 1 &  \\
                                        & 1x/year &  2 &\\
                                        & 1x/month &  3 & \\
                                        & 1x/week & 4 & \\
                                        & 1x/day &  5  &\\
            \midrule 
               Pain Factor  & 1-5 & 1-5 
                            & The severity of the stress should already be measured on a scale of 1-5  \\
            \midrule 
               Quality attributes  & n  & 1-5 
                            & Number \textit{n} of quality attributes affected in addition to maintainability   \\
            \midrule 
               Contagiousness   & decreases &  0  & \\
                            & stagnates &  3 &\\
                            & increases &   5 &\\
            \bottomrule
        \end{tabular}
        \caption{Integer assignment to attributes for a mean value calculation}
            \label{tab:meanIntegers}
    \end{table*}
    Contagion must be considered separately, as it is not a linear change. If repayment costs decrease over time, the overall priority should also decrease, which is achieved by setting a value of 0. If repayment costs remain the same, the value should not affect the mean values. However, this is difficult to achieve mathematically because the calculated mean value can vary from issue to issue. The values given here can therefore only be used to create an approximation. The team is free to flexibly adjust the integers of the contagion to the situation when determining each mean value.

    The repayment costs are not taken into account here (in contrast to the ROI variant). By means of a ``low-hanging fruit'' visualization (see~\Cref{fig:VisualisierungPlanung}), the combination of these two values can lead to a sequence for repaying individual technical debt items.

    The advantage of this method is that the calculation can be done in your head and is therefore more pragmatic and reliable than an ``educated guess.''
    
    \textit{\textbf{Return on Invest (ROI):}}
    The ROI is calculated using the interest rate, the probability of interest, and the effort in person-days. The result is the number of months it takes for the effort to repay the technical debt to break even.
    To calculate the ROI, we set the following values for the combo box fields:
    \begin{itemize}
        \item Interest in minutes: 15; 60; 240; 480; 960
        \item Interest probability in ``per month'': 30; 4.5; 1; 0.0027; 0.0013
        \item Expense in PD: as specified
    \end{itemize}
    From this, we calculate the monthly interest burden, as specified in~\Cref{eq:interestburden}, and the ROI in months, as specified in~\Cref{eq:roi}.
    
    \begin{equation} 
        \text{\textit{interest burden}} \enspace \text{minutes/month} = \text{\textit{interest}} \enspace \text{minutes}  * \text{\textit{interest probability}} \enspace \frac{1}{\text{month}}
    \label{eq:interestburden}
    \end{equation}
    
    \begin{equation}
      \text{\textit{ROI}} \enspace \text{months}= \frac{\text{\textit{effort}}  \enspace \text{minutes}}{\text{\textit{interest burden}} \enspace \text{minutes/month}}
    \label{eq:roi}
    \end{equation}
    \vspace{0.3cm}
    
    The ROI can then be used to derive priorities in various ways:
    \begin{itemize}
        \item The ROI serves as additional information for an ``educated guess'' priority. This means that not only does the ROI determine the final priority, but other factors (e.g., additional quality attributes affected; see section below) are also taken into account in the determination.
        \item The ROI is directly transferred to a priority. Priorities are about ratios, i.e., the question of whether one technical debt is more important than another. Therefore, based on the determined ROI, it can be distributed as evenly as possible among the existing priorities. This distribution may vary from company to company. An example is shown in~\Cref{tab:roiPrio}.
    \end{itemize}

    \begin{table*}[htbp]
        \centering
        \begin{tabular}{lcl}
            \toprule
            ROI & Prio ID & Priority \\ 
            \midrule 
                $<$ 1 month   & 5 & very high \\
                $<$ 2 months  & 4 & high       \\
                $<$ 1 year    & 3 & medium        \\
                $<$ 3 years   & 2 & low        \\
                $\geq$  3 years  & 1 & very low   \\
            \bottomrule
        \end{tabular}
        \caption{Example of determining priority based on return on investment (ROI). The actual distribution may vary depending on the team/company.} 
            \label{tab:roiPrio}
    \end{table*}

    The advantage of ROI is that it presents a profit that is easily understood by management. ROI is based on the same principles as project planning as a whole, i.e., it is based on expert estimates. However, this also means that it is subject to the same inaccuracies. The team should decide together with management whether accuracy is desired and make it clear that the estimates (as in project planning) may be subject to errors.

    \paragraph{Resubmission}
    
    In an agile environment, there are generally two types of prioritization. 
    On the one hand, each issue is assigned a priority. On the other hand, issues in the backlog can be moved up and down, i.e., the order of the issues in the backlog specifies a further—usually more relevant—priority.
    Issues related to technical debt, as well as some other issues, are repeatedly prioritized downwards and eventually end up at the bottom of the backlog. If the number of such issues becomes excessive, they are no longer reviewed.
    The following procedure has therefore become established:
    
    \begin{myframe}
        Each technical debt issue is assigned a resubmission date.
        In each refinement, the list of all technical debts is then sorted by resubmission date, and all issues with an expired resubmission date are reevaluated.
    \end{myframe}

    To avoid starting from scratch every time a reassessment is carried out, it is essential that all relevant information and decisions are recorded in writing in the issue.
    In particular, it should be noted, for example, in a comment, if a specific situation is anticipated that will lead to a change in the assessment. Examples: ``Reassessment necessary after business decision <xy>'' or ``Issue should be obsolete after system <xy> is shut down.''

    A resubmission date has the following advantages:

    \begin{itemize}
        \item Issues are not permanently overlooked at the end of the backlog (as explained above).
        \item If the priority depends on a business decision that has not yet been made, the resubmission date can be set to the estimated date of this decision in order to reevaluate the issue.
        \item If a change is planned that will render the issue obsolete in the near future (e.g., switching on or migrating a system), the date can be set to the date of the expected shutdown. On the resubmission date, a decision will be made regarding whether the issue can be closed.
        \item If the priority of the issue is low and it makes sense not to repay the technical debt but to pay the interest permanently, the resubmission date can be set. On the date, it will be checked whether repayment has now become sensible due to interim changes. The issue thus remains open and can also be used for risk assessment (see~\Cref{sec:repay}) without being repaid.   
    \end{itemize}

    \paragraph{Quality attributes}
    As already mentioned in~\Cref{tab:attributesII}, quality attributes can be used for prioritization. These can be used, for example, to overlay a calculated priority based on interest rates, interest probability, and effort, e.g., when security risks need to be minimized.
    
    Depending on the context (system/industry/company), different quality attributes are most important. For example, security is most important in the context of banks, whereas performance (efficiency) and safety may play a greater role in the embedded sector.

    \begin{myframe}
        Each team should discuss which quality attributes are, in general, most relevant to their own system in their own context/company. To do this, the quality attributes should be put in order. 
        The aim is not to establish a fixed and universally valid order, but to discuss the attributes and develop a common understanding of which attributes are relevant to their own system.
        The relevant quality attributes can then be used to influence decision-making/prioritization.
    \end{myframe}

    In~\Cref{tab:qas}, we briefly introduce all quality attributes according to ISO-25010 and explain them in simple terms using our own words and example questions.
    Those who are already familiar with ISO quality attributes are welcome to skip this section.
    
    In addition to the ISO quality attributes, other quality attributes (e.g., ``sustainability'') or the sub-categories of the ISO standard (e.g., ``installability'' as part of ``flexibility'') may be useful and can be applied. It is also perfectly legitimate to use old terms, such as ``performance'' as an old term for (and component of) ``performance efficiency'', or ``usability'' instead of ``interaction capability''. 
    The only important thing here is that there is clarity and agreement within the team about the meaning of the terms.

    In practice, maintainability, security, performance, and reliability are often influenced by technical debt, whereas the other quality attributes are less frequently impacted.
    
       	\begin{sidewaystable*} [!htbp]
    	    \centering
        	\begin{tabular}{p{2.3cm}p{5cm}p{14cm}}
    			\toprule
        	   	Quality attributes  & Sub-Categories & Explanation   \\ 
    			\midrule 
    			Functional Suitability 	
                        & F. completeness; F. appropriateness; F. correctness  
                        This concerns the appropriateness of the chosen solution, for example, the question: Does the solution truly meet the users' needs? From a technical debt perspective, this also includes, for example, systems/system components that are no longer in use and should be shut down to save on maintenance costs. \\ 
    			\midrule 
    			Performance Efficiency  	
                        &  Time Behaviour; Ressource utilization; Capacity 
                        & This is about the efficiency of the solution. Does the chosen solution perform well? How many resources (e.g., memory) are required? The last point now also encompasses issues of sustainability in terms of environmental and climate concerns. \\ 
    			\midrule 
    			Compatibility  	
                        & Co-Existence; Interoperability 
                        & This concerns the interaction of different systems. Can the chosen solution be integrated into the existing system landscape? \\ 
    			\midrule 
    			Interaction Capability  	
                        & z.B.: Learnability, User error protection; Inclusivity; u.v.m.   
                        & Before 2023, this area was called ``usability,'' and this term is usually easier to understand. It refers to the ease of use for the user. Is the system easy to learn? Is the user protected from user errors, e.g., through warning messages? Can the system also be used by people who are blind/deaf? \\ 
    			\midrule 
    			Reliability  	
                        &  Faultlessness; Availability; Fault Tolerance; Recoverability 
                        & This concerns errors that can occur in a system. How often/how many errors occur? Are there system crashes? How long must the system be available per year? \\ 
    			\midrule 
    			Security  	
                        &  z.B.: Confidentiality; Integrity; Authenticity u.v.m.  
                        & This is about protecting the system against misuse. Is the data encrypted? Is it possible to trace who changed which data? \\ 
    			\midrule 
    			Maintainability  	
                        & Modularity; Reusability; Analyzability; Modifiability; Testability 
                        & This is the real core of technical debt. How easy is it to change the system? Is the system modular? Can the code be easily read by developers who are not familiar with it? (Note: Testability does not refer to the existence of tests, but rather whether the code is written in such a way that it can be easily tested.) \\ 
    			\midrule 
    			Flexibility  	
                        & Adaptability; Scalability; Installability; Replaceability  
                        & The question here is whether the system can be easily adapted to other circumstances. How easily can peaks in demand be handled? How easily can the database be replaced? \\ 
    			\midrule 
    			Safety  	
                        & z.B.: Risk Identification; Fail Safe; Hazard Warning u.v.m.  
                        & This concerns safety when using the system. Could misuse result in injury to a person (e.g., in mechanical engineering)? Are users warned before they endanger the environment, including the people in it (e.g., nuclear power plants, MRI)? \\ 
     			\bottomrule
    		\end{tabular}
    		\caption{Quality attributes according to ISO-25010~\cite{iso_25010_2023}}
                \label{tab:qas}
    	\end{sidewaystable*}

\newpage

\subsubsection{Repay Technical Debt}
\label{sec:repay}

    There are various approaches to repaying technical debt, each of which is appropriate in different situations.
    Situational repayment has therefore proven to be best practice. This means that there is no single perfect method for repaying technical debt; rather, different approaches are appropriate in different situations.
    \Cref{tab:repay} provides an overview of the various repayment methods and the situations in which each method is appropriate (use case). 
    The repayment methods can be divided into three categories: ignore, rewrite, and refactor.
    For the sake of completeness, we include ``Ignore'' as a repayment method, even though no explicit repayment takes place and the term ``repayment'' may be misleading in this context.

    \begin{myframe}
        Each team should decide for itself which methods it wants to use.
        In particular, a decision should be made regarding whether to use a quota method (e.g., 15
    \end{myframe}
    
    If a team is under a lot of pressure from customers and new customer requests and/or has a very confusing backlog with many ``old'' issues, it has proven useful to work with a quota solution. 
    Teams that can work more flexibly and independently have generally not found the quota solution necessary.
    The decision on whether a quota solution is suitable should be made by the developers. The scope may be negotiated with the product owner or management.

            \begin{sidewaystable*} [!htbp]
    	    \centering
        	\begin{tabular}{p{1cm}p{3cm}p{9.5cm}p{8.5cm}}
    			\toprule
        	   	Cate- & method  & Explanation & Usage-Scenario   \\ 
        	   	gory &   &  &    \\ 
    			\midrule 
                 \multirow{2}{*}{\raisebox{-1,9\height}{\rotatebox[origin=c]{90}{IGNORE}}}
    			 & Impediment/ Road-Block 
                        & No one cares about technical debt. The debt is not reduced. Interest is paid (unconsciously) on an ongoing basis. Repayment only occurs if further development is not possible without repayment. The risk that further development will no longer be possible is (unconsciously) accepted.
                        & A purely functional/customer-driven development; no awareness of and no management of technical debt.       \\ 
    			 & Pay interest
                        & The debt is not being reduced. Interest is being paid on an ongoing basis (or until the next resubmission date).
                        & A Repayment is not reasonable for economic or other reasons, such as affected quality attributes or developer frustration. \\ 
    			\midrule 
    			\multirow{1}{*}{\raisebox{-1,0\height} {\rotatebox[origin=c]{90}{REWRITE}}} 
                & Magic 
                        & Technical debts ``disappear'' without explicit repayment because systems are rewritten, replaced by a purchased system, or simply shut down. 
                        & Rewriting when using very outdated technologies or as part of planned migrations (e.g., cloud migration); shutting down systems/parts of systems that are no longer used by users; replacing self-developed systems with market provider systems for economic reasons.   \\ 
    			\midrule 
                 \multirow{4}{*}{\raisebox{-3,5\height}{\rotatebox[origin=c]{90}{REFACTOR}}} 
              & Calculate/prove benefits
                        & Every technical debt must ``prove'' its usefulness, e.g., because an ROI will be achieved in a few months. Repayment of ``low-hanging fruit,'' i.e., issues with high priority and low effort.
                        & The product owner should schedule these issues in the list of all issues for a sprint because their benefits are obvious. Schedule them when there are only a few SPs left to allocate in the planning, or when an issue in the sprint was completed faster than expected.    \\ 
    			 & Contingent 
                        & Setting a quota for repaying technical debt: x\% of a sprint's SPs; every xth sprint solely for repayment; project for repaying technical debt (``quality initiatives'')
                        &  For teams that are generally under a lot of pressure from customers and new customer requests, to ensure regular repayment and maintenance.   \\ 
    			 & Evolution-based
                        & Repayment of technical debt in an (architectural) component that is being changed as part of an epic/project/feature. Goals: (1) Minimize the risk of technical debt causing disruption. (2) Leverage synergies because the relevant section is being changed and tested anyway.
                        & When a major change to a specific component is planned.         \\ 
    			 & Polluter pays principle
                        & Repayment of technical debt incurred as part of a major change (e.g., epic, project, feature development). Create an issue for each piece of technical debt incurred and repay it after the change has been deployed.
                        & When a major change must be completed by a certain date and technical debt is incurred in order to meet that deadline.          \\ 
     			\bottomrule
    		\end{tabular}
    		\caption{Repayment methods and the associated application scenarios}
                \label{tab:repay}
    	\end{sidewaystable*}

\newpage
\subsubsection{Visualize Technical Debt in the Backlog}
\label{sec:visualize}

    Technical debt is visualized based on its issues and the estimated values for the attributes (see~\Crefrange{tab:attributesI}{tab:attributesII}).

    The visualization has two main objectives:
    \begin{itemize}
        \item Support for planning, i.e., prioritization and monitoring of technical debt
        \item Raising awareness of technical debt, especially among non-technical stakeholders such as customers, business management, project managers, etc.
    \end{itemize}

    \begin{myframe}
        Interactive visualizations (low-hanging fruits, bar charts by component or quality attributes) have become particularly popular for planning purposes, as they filter an additional list of technical debt issues.
        This list should display the following attributes and be sortable by all attributes:
        ID, title, priority, contagion potential, interest, interest probability, interest burden, ROI, resubmission date, link to the original issue.
    \end{myframe}
    
    \begin{figure*}[!htbp]
        \centering
        \includegraphics[width=\textwidth]{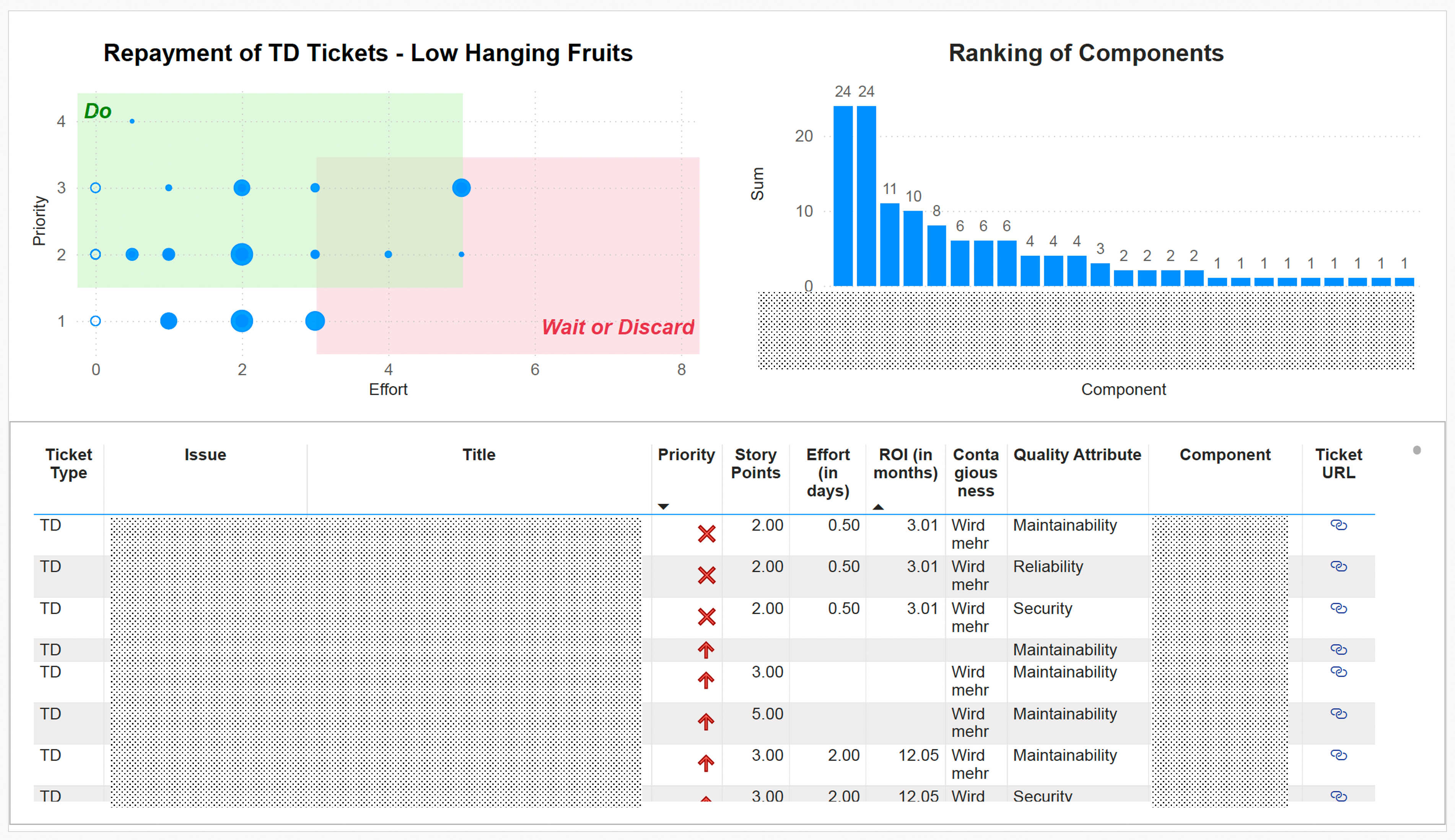}
        \caption{Example of a visualization of technical debt issues for planning purposes (anonymized)}
        \label{fig:VisualisierungPlanung}
    \end{figure*}
    
    The list is controlled via a graphical visualization, whereby the following three visualizations (sorted by importance) have become established for this purpose:
    \begin{enumerate}
            \item Low-hanging fruits (\Cref{fig:VisualisierungPlanung}): In a coordinate system with ``effort'' on the x-axis and ``priority'' on the y-axis, points indicate technical debt issues. The size of the points symbolizes the number of issues at that point. The list displays the technical debt associated with the point clicked by the user.
            \item Components (\Cref{fig:VisualisierungPlanung}): Issues are displayed per component in a bar chart or pie chart. The list displays the technical debt associated with the component clicked by the user.
            \item Quality attributes: Issues are displayed per quality attribute in a bar chart or pie chart. The list displays the technical debt associated with the quality attribute clicked by the user.
    \end{enumerate}

    \begin{figure*}[!htbp]
        \centering
        \includegraphics[width=\textwidth]{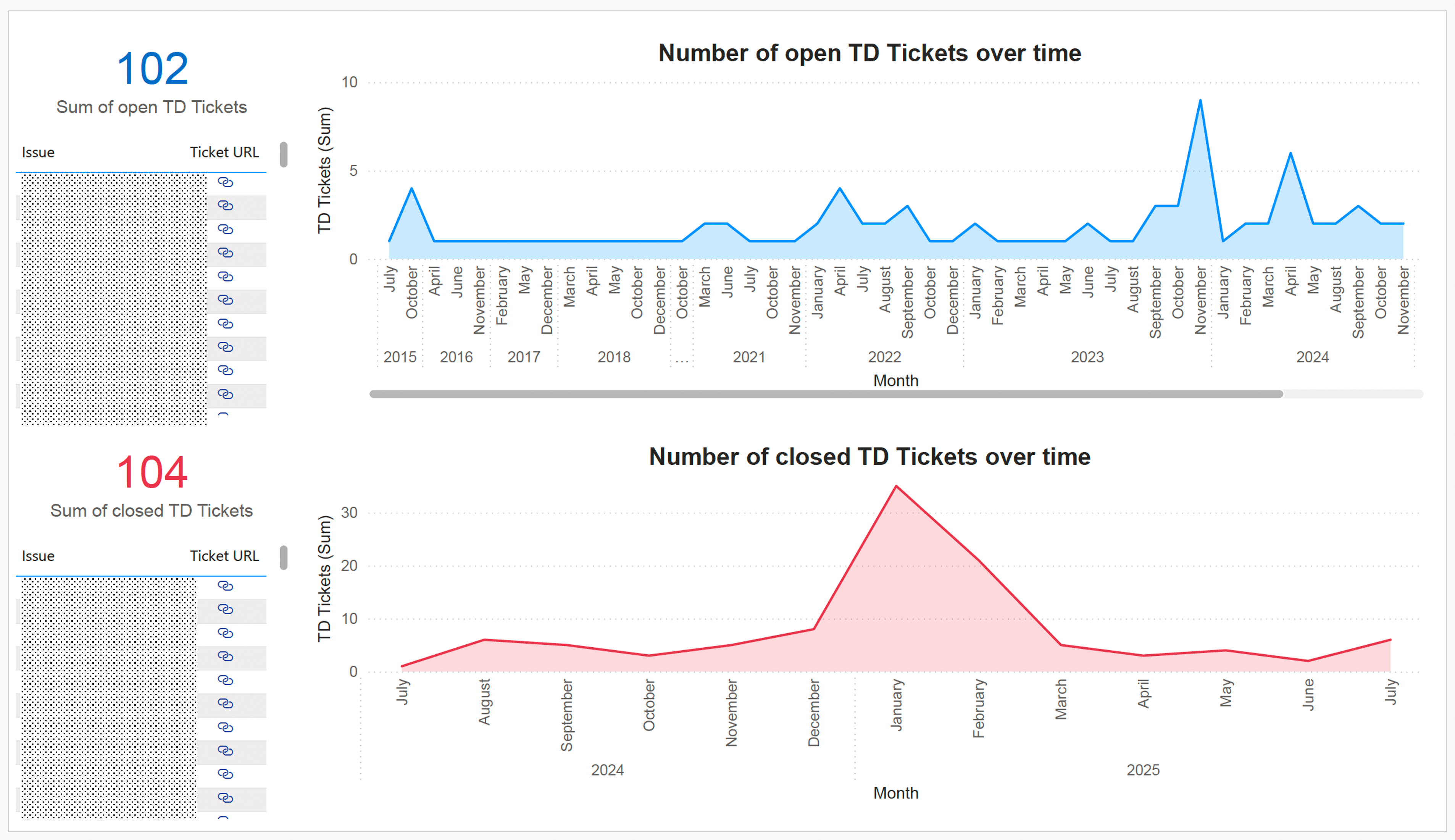}
        \caption{Example of a visualization of technical debt issues for management and monitoring purposes (anonymized)}
        \label{fig:VisualisierungManagement}
    \end{figure*}
    
\begin{figure*}[!htbp]
        \centering
        \includegraphics[width=0.9\textwidth]{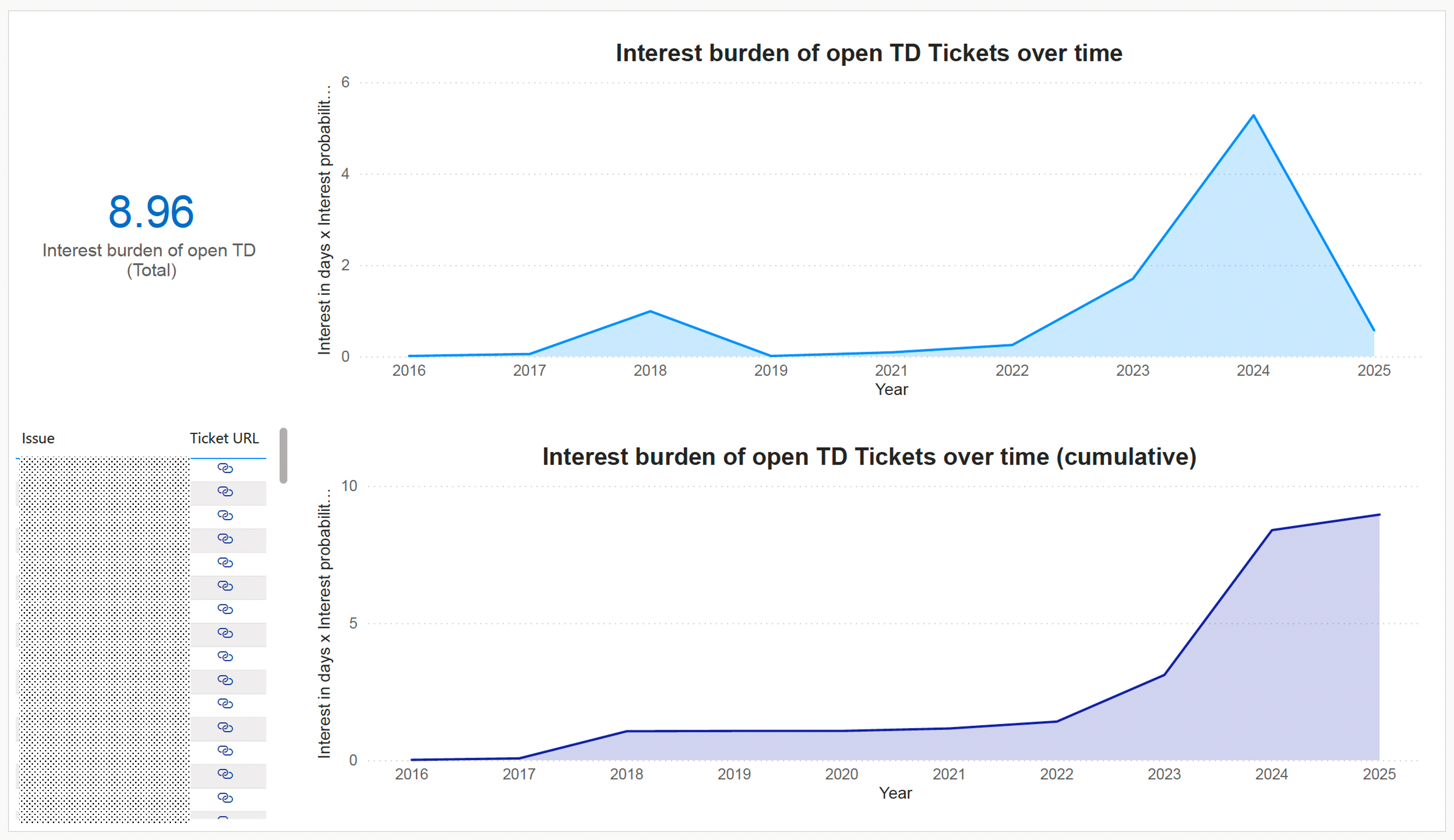}
        \caption{Example of a MISLEADING visualization of the interest burden, but only for currently open tickets and their ``open'' date, and not as a time series (anonymized)}
        \label{fig:VisualisierungInterestburden}
    \end{figure*}

    The visualizations used to raise awareness of technical debt also serve to monitor the management process and are primarily based on time series:
    
    \begin{itemize}
        \item Number of open technical debt issues per month (\Cref{fig:VisualisierungManagement})
        \item Number of closed technical debt issues per month (\Cref{fig:VisualisierungManagement})
        \item Number of open technical debt issues per month
        \item Interest burden (see~\Cref{eq:interestburden}) of all open technical debt issues per month (\Cref{fig:VisualisierungInterestburden})
    \end{itemize}

    \paragraph{Technical obstacles}

    Most issue tracking systems offer insufficiently flexible visualization options. 
    It is therefore advisable, as already mentioned in~\Cref{sec:tool}, to connect the issue tracker to a reporting tool such as MS Power BI or Tableau.
    This creates development and maintenance problems. However, a bigger problem is that user acceptance of the visualization decreases when a separate and previously unknown tool has to be used.

    In addition, it is particularly challenging to determine the number of open technical debt issues per month. Issue tracking systems store the ``open'' and ``closed'' dates. To determine the number of open issues for each month, a code snippet must be used to iterate through all months in a loop and then check for each month and each issue whether it was open. 
    A counter is incremented for the number of open issues. For the interest burden of open issues, the interest burden must be determined and totaled for each open ticket.
    The code snippet to be developed makes development and maintenance even more difficult.
    A visualization of the interest burden without this calculation is misleading, as the interest burden always appears to be increasing because issues that have already been closed are no longer factored into the calculation.

\newpage

\subsection{Getting Started}
\label{sec:gettingstarted}

    To start the process, a few steps are necessary, which are listed here in chronological order:

    \begin{enumerate}
        \item The ``technical debt manager'' (TS manager; see~\Cref{sec:orga}) is appointed in a joint meeting. From then on, they control the further steps in accordance with these instructions.
        \item A keyword name for technical debt (e.g., ``TD'') and issues that have already been searched but not identified as technical debt (e.g., ``Non-TD'') is defined.
        \item All existing issues are reviewed, and it is determined whether or not they constitute technical debt. The issues are tagged accordingly (``TS'' / ``Non-TS''). See also~\Cref{sec:questions}. (``Non-TS'' is useful for performing this analysis in several smaller blocks so that you know which issues have already been considered.)
        \item All team members record all technical debt they have in mind but have not yet documented, and tag it as technical debt.
        \item Discuss within the team which of the attributes from~\Crefrange{tab:attributesI}{tab:attributesII} are useful for this team/company.
        \item Create a new issue type for technical debt with the selected attributes. Examples can be found in the appendix~\Crefrange{fig:Ados}{fig:GitLab}
        \item Transfer all issues tagged as technical debt to the new issue type
        \item Evaluate all technical debt issues. There are two options for this:
            \begin{itemize}
                \item Specific meetings are scheduled for all team members to refine technical debt issues. These issues are discussed and evaluated one by one, meaning the attributes are filled in. This approach is especially recommended at the beginning of the evaluations so that the team can develop a shared understanding of how to evaluate the attributes.
                \item Each team member takes on 3-4 technical debt issues per week and evaluates them individually ``in advance.'' In particular, the team member sets the resubmission date and can thus determine how urgently the issue needs to be considered by all team members. If the team member is unsure about their assessment, they can, for example, choose a short-term resubmission date. If the issue is obviously less relevant, the resubmission date can be set for a far future. During refinement, when all issues with expired resubmission dates are reviewed, the issue is then evaluated jointly by all team members. This distributes the burden of evaluation over time, eliminating the need for separate meetings.
            \end{itemize}
        \item The team decides (in a separate meeting, if necessary) which repayment procedure should be used (in which situations).
        \item The TS manager ensures that the repayment agreements are adhered to.
        \item The TS manager clarifies which visualization tools can be used, links the issue tracker to the visualization tool, and creates initial visualizations.
        \item The visualizations are used by the team and adjusted by the TS manager as needed.
        \item In all other issue types, the attributes for avoiding technical debt are added in accordance with~\Crefrange{tab:EntscheidungenBewusst}{tab:EntscheidungenKlug}. This can be done as an attribute field or as part of a DoR.
        \item For all issues (including technical debt issues!) that are now being handled in refinement, the attributes are evaluated.
    \end{enumerate}

\newpage
\section{Typical Challenges in Managing Technical Debt}
\label{sec:problems}

During the introduction of technical debt management, typical mistakes (\Cref{sec:mistakes}) and questions (\Cref{sec:questions}) arise time and again. The following two chapters provide an overview of these topics and suggest solutions. In particular, the typical mistakes from~\Cref{sec:mistakes} should be taken into account in order to successfully introduce the process. 

\subsection{Typical Mistakes}
\label{sec:mistakes}

    There are mistakes that teams repeatedly make during implementation. These ``mistakes'' are usually caused by the fact that the issue of technical debt is overlooked in the hectic day-to-day business.
    The role of the TS manager helps to keep the topic in mind during day-to-day business. 
    However, mistakes can still occur, despite having a TS manager, and the following list can be passed on to future TS managers as a learning experience.
    
    \subsubsection{We all remember it.}        
        There are always situations in which a task in the process is overlooked, for example, discussing drawbacks and risks to avoid technical debt (see below).
        This is often because there is no clear responsibility for this task. 
        The team decides to remember it (``we'll all do it''), but in the end, no one takes responsibility for the task, and it is forgotten. This phenomenon particularly affects the management of technical debt, as in this case, by definition, these are issues that are not (immediately visible) business-critical.

        \textbf{Solution:}
        You need someone who will repeatedly remind you of the chosen process and the resulting tasks (e.g., discussing drawbacks and risks). In the initial phase of establishing the process, it has proven useful for this person not to be part of the team. This allows team members time to become familiar with the process and integrate it into their daily routines. 
        In the long term, the TS manager should then take over the task. Other people can also take on tasks, but even then, there should be clear accountability. 
    
    \subsubsection{The attributes all make sense -- we use them all.}
        All attributes listed in the~\Crefrange{tab:EntscheidungenBewusst}{tab:attributesII} initially appear useful and important, which is why they are listed here. However, this creates a significant risk of using all attributes in your own template, thereby making the process (unnecessarily) complicated.

        \textbf{Solution:} 
        Only the ``best practices'' should be used for the template initially. A maximum of one or two attributes from the ``nice-to-haves'' should be used at the beginning, but preferably none at all. During daily work, i.e., the application of the process, it will then become apparent whether additional attributes are helpful and should be added.
        
    \subsubsection{Forgetting the new attributes for non-technical debt issues}
        When introducing the new process, the focus is typically on managing existing technical debt. Avoiding new technical debt is often seen as secondary. As a result, people often forget to fill in the ``attributes for all issue types'' (see~\Cref{sec:prevent}, \Crefrange{tab:EntscheidungenBewusst}{tab:EntscheidungenKlug}). 
        
        \textbf{Solution:}
        It is advisable to appoint someone from the team to closely monitor this. This could also be the TS manager. And no, ``all of us'' will not pay attention to this (see above). Responsibility must be assigned to ensure that the issue is successfully addressed.

    \subsubsection{Alternatives? There aren't any.}
        In everyday working life, it's easy to overlook the attribute ``alternative.'' 
        The motto is: briefly mentioned and then forgotten again. ``Alternatives? Nope, I can't think of any. So there aren't any—check.'' In many cases, this is the pragmatic approach; however, you must be careful that the approach doesn't take on a life of its own. 
        
        \textbf{Solution:} How to find alternatives is described in detail in~\Cref{sec:questions}.
        
    \subsubsection{Unconsciously accumulating technical debt: Planning Poker is just annoying}
        In many teams, planning poker is no longer used or is used only half-heartedly. The effort seems too small for the benefit, and ``we all know what the right number is anyway.'' In some teams, it even goes so far that the product owner makes the estimates or at least gives the first estimate. 
        
        The problem with this is the anchoring effect, i.e., the tendency to prefer what is said or heard first. This is a cognitive bias, i.e., a mistake our brain makes for efficiency reasons, and one that is difficult for us to escape. Once a value has been put forward (e.g., 2 SP), it is difficult to name a completely different value (e.g., 8 SP). This leads to misestimations and thus to time pressure. However, time pressure is one of the main causes of conscious and unconscious technical debt. In particular, if the developer feels a moral responsibility because they accepted the estimate, they are more likely to want to stick to the estimate -- at the expense of quality, if necessary. 

        \textbf{Solution:}
        The best thing to do is actually to resume Planning Poker. A lightweight variant is to at least provide a clear justification for each estimate. This allows another person to refer to the reasoning if they want to suggest a completely different value, e.g., ``You only estimate 2 SP, but you forgot that there is also the dependency on the <xy> system, so I see the effort as significantly higher.'' 
        
        Under no circumstances should the product owner give an estimate. Estimates are always and exclusively made by the experts concerned.
        
    \subsubsection{Unconsciously taking on technical debt: But that has to be included in the sprint now.}
        Unfortunately, there are often situations where issues must be included in the sprint because they have a short deadline or have already been ``promised'' to the customer. However, if uncertainties still surround these issues, e.g., feedback from tool providers or other teams regarding the details of implementation, the issue may not be implemented optimally because important information is missing. Out of a misguided desire to accommodate the customer/requester, a poorer solution is implemented, thereby unknowingly incurring technical debt.
        
        \textbf{Solution:}
        An issue may only be included in the sprint once all information is available. Here, the product owner has a duty to communicate openly with the customer/requester. Often, the customer/requester is not even aware that a less-than-optimal solution is being implemented because communication with the customer/requester is avoided.
        It should be made clear to the customer/requester that the deadline can only be met at the expense of 
        quality and what specific consequences the lack of quality can have in the long term (see~\Cref{sec:causesconsequences}).

        A similar communication problem sometimes arises between developers and product owners. Developers should also make it clear to the product owner that the deadline can only be met at the expense of 
        quality and what specific consequences can be expected.

    \subsubsection{Consequence of technical debt: Poor maintainability}
        The attribute ``risks of non-repayment,'' i.e., the consequences of technical debt, invites us to write something like ``poor maintainability,'' which has indeed often happened in practice. However, this is already the definition of technical debt, or at least part of it.
        The consequences should be described in more detail and understood by non-technical stakeholders.
        
        \textbf{Solution:}
        There is a method called the ``5 Whys''~\cite{serrat_five_2017}, which involves identifying the root cause of a problem by asking multiple (five) questions about the cause (``Why'') in succession. This method can be adapted for this purpose by repeatedly asking why the stated issue is a problem, e.g.: 
        (1)~Why is it a problem if maintainability is poor here? -- Because changes take longer and are risky. 
        (2)~Why is it a problem if changes take longer and are risky? -- Because we change this component frequently. 
        (3)~Why is it a problem if we change this component frequently? -- Because we will no longer be able to meet deadlines in the future, and the risk of introducing bugs will increase.
        (4)~Why is it a problem if we don't meet schedules and introduce bugs? -- Because customers will be dissatisfied.  
        (5)~Why is it a problem if customers are dissatisfied? -- The reputation of the system/company will decline.
        The point here is not to record all the details, but to extract the essential information and write that down. In the example, that would be: ``Changes to this central component are frequent. These changes will take longer in the future and carry an increased risk of side effects, such as bugs or performance losses, which could upset our customers.''
        
    \subsubsection{The components ``Test,'' ``Documentation,'' and ``Logging''}
        Some teams find it difficult to list the (architectural) components. In particular, so-called ``cross-cutting concerns'' (i.e., tasks that span multiple components), such as testing, documentation, logging, monitoring, feature flag removal, and even refactoring, are often recorded as components.
        Component assignment is important in technical debt management, especially for evolution-based repayment. The goal is to identify the technical debt items contained in the component that is currently being modified to meet a functional requirement. However, if a cross-cutting concern, such as ``documentation,'' has been selected as a component, this does not provide such information and is therefore meaningless. It is not clear which component is inadequately documented. It is unclear whether there might be faulty documentation of the component to be changed, and whether the documentation should be optimized along with the change in order to exploit synergies.
        
        \textbf{Solution:} 
        When recording the components, a counter-test should always be performed to determine whether it is a ``cross-cutting concern.'' To do this, consider whether there could be a functional change associated with this component. 
    
    \subsubsection{Nothing is 'very'}
        If a 5-point scale from ``very low'' to ``very high'' is chosen for the values in a combo box, there is a risk of ``central tendency bias,'' i.e., the unconscious tendency of the human brain to choose a middle value. Extreme values, on the other hand, are avoided. As a result, a theoretical 5-point scale quickly becomes a practical 3-point scale.
        
        \textbf{Solution:} 
        In addition to the values from ``very low'' to ``very high,'' there should be a clarification of what is meant by each value. We have already shown this using the example of interest rates and interest rate probabilities in~\Cref{tab:attributesI}. There, specific values such as ``$<$ 15 min.'' or ``$>$ 1x/year'' are also given, which gives developers a concrete idea of the values and allows them to use ``very low'' and ``very high'' normally.

    \subsubsection{Resubmission: As a precaution, next week}
        A common mistake is to set the resubmission date too soon. 
        
        \textbf{Solution:}
        When setting the resubmission date, it is not about when you want the issue to be resolved, as this is determined by the priority.
        It involves waiting for specific events (e.g., management decisions, developers returning from vacation) or observing whether general conditions have changed unexpectedly in other ways (e.g., changes in management, new customers, new technologies on the market). Therefore, it makes no sense to assign a resubmission date in a few months if no specific decision is to be awaited. In these cases, the date should be at least 6 months (preferably more) in the future, as an essential change in the general conditions is unlikely during this time.
    
\subsection{Typical Questions}
\label{sec:questions}

    Some frequently asked questions in practice will be addressed here. It makes sense to read the specific answers to questions when they arise.

    \subsubsection{Too many issues in the backlog—how do I get started?}
        At the beginning of the process, all issues in the backlog must be searched, and technical debt must be marked. This can be a tedious task for large backlogs with long-standing issues.
        
        \textbf{Solution:}
        An initial assessment is best made intuitively, based only on the title and the indicators mentioned in~\Cref{sec:TDindicators}. The label for technical debt can be assigned generously, i.e., better once too often than once too little. Since all issues are discussed again in refinement, it is still possible to decide that an issue does not describe technical debt after all.
        In cases of very large backlogs with very old issues, you can also start by focusing on issues created within the last year (or the last two years). Older issues may still contain further technical debt, but these are usually either less problematic or they reappear and can then be recorded as new issues. In this case, it is more important to start the process than to have all issues fully recorded.
                
    \subsubsection{How can I find alternatives?}
        Some teams struggle to identify alternatives to the solution they initially proposed. Cognitive biases unconsciously lead them to consider only one solution.

        \textbf{Solution:}
        Teams often start by gathering all the alternatives and then comparing their pros and cons. However, it has turned out that using the discussion order shown in~\Cref{tab:EntscheidungenKlug} makes sense. First, the quality attributes affected by the initially proposed solution should be discussed. Based on this, the drawbacks and risks can then be derived (e.g., if performance is negatively affected). Starting from the drawbacks and risks, it is then easier to identify an alternative solution that may not entail these drawbacks/risks.
        
    \subsubsection{Can I use other terms?}
        The terms used here for technical debt are based on the metaphor of financial debt. This metaphor was originally developed to make the topic of ``technical debt'' more understandable to business stakeholders~\cite{Cunningham1992}. Some of the terms are based on other scientific publications, such as the concept of contagion~\cite{Martini2015a}.
        These terms are sometimes difficult for development teams to understand or comprehend. The terms have also not become as established among business stakeholders as originally hoped~\cite{Schmid2013, wiese_it_2023}.
        
        \textbf{Solution:}
        It is therefore perfectly legitimate to use other terms as long as their meaning is clear to everyone involved. For example, some teams have adopted the term "interest frequency" instead of "interest probability" because it makes the assignment of combo box values easier to understand.
    
    \subsubsection{There are several consequences. How do I assess this?}
        As already mentioned in~\Cref{tab:attributesI}, there can be several consequences of technical debt, all of which should be recorded in the attribute ``Risks of non-repayment.'' However, there are also several different ``interest charges,'' but only one attribute ‘Interest’ and one attribute ``Interest probability.'' 
        Theoretically, the interest rates for all consequences would have to be estimated with their respective interest rate probabilities and then summed, which is neither cognitively nor computationally feasible.
        
        \textbf{Solution:} 
        A pragmatic approach that has proven successful is to evaluate the most serious consequences.
        The team could also decide to set interest and interest probability higher than appropriate for the most serious consequence in order to account for further consequences. However, this approach also requires increased cognitive effort and should only be used if the team has already gained some experience with the process. 

    \subsubsection{How do I handle ongoing or recurring maintenance tasks such as updates?}
        Some maintenance tasks need to be performed on an ongoing basis, such as updating 3\textsuperscript{rd}-party libraries or tools used. In these cases, it would be too time-consuming to create an issue for every version change; instead, some teams maintain a simple list of tools with their versions and do not want to deviate from this approach.
        
        \textbf{Solution:}
        In this case, you can create a single 'update' issue and describe and evaluate the most urgent update within it. As soon as the priority of this most important update is higher than that of all other technical debt issues, the issue has to be implemented, i.e., the update has been carried out. After that, an issue is created for the next most important update. 
        
        Of course, the team only makes a rough estimate of which update is most important, rather than evaluating the priority as precisely as in the technical debt issues. This inaccuracy is a pragmatic compromise. If there are disagreements about which update is the most urgent, two or more update issues can be created and evaluated.

\newpage
\section{Outlook and Scaling of Technical Debt Management}
\label{sec:conclusion}


    The introduction of technical debt management described above serves individual teams. In a corporate environment, it makes sense to consider whether and how this approach can be scaled, i.e., applied to many or all teams within a company.

    Using the process described here across all teams within a company offers many advantages.
    There are benefits for each individual team in their daily work:
    \begin{itemize}
        \item Overview of technical debt within the team
        \item Assessment of risks arising from technical debt 
        \item Prioritization of various technical debt issues 
        \item Clearly defined and consistent repayment of technical debt
        \item Ability to provide information and arguments to management 
    \end{itemize}
    There are further advantages for management when the process is used by several or all teams and the resulting data is reported in summary form:
    \begin{itemize}
        \item Overview of technical debt per team and per component, enabling the identification of ``hot spots.''
        \item Assessment of the risks arising from technical debt in cross-team projects
        \item Prioritization of support for teams, e.g., relieving the team/component of customer requests in order to have time to repay technical debt or increase the team's staffing levels
        \item Clearly defined and therefore calculable procedure for repaying technical debt
    \end{itemize}

    \begin{myframe}
        For scaling, a team within the company should first be given the time to thoroughly address the topic and establish a process based on these guidelines. The result will be an issue type and a corresponding visualization. Both can then be used by other teams with significantly less initial effort.
        Nevertheless, these other teams must also be given time for the initial discussion and evaluation of technical debt issues.
    \end{myframe}
    
    According to initial estimates, the effort required for the first team is 5-6\% of working time in the initial phase and 3\% of working time on an ongoing basis for the process itself.

    In our view, there are two possible ways to encourage other teams to use this system:
    \begin{enumerate}
        \item Dissemination of the process ``from the top down,'' e.g., via management or architecture committees, i.e., the team architects.
        \item Dissemination of the process ``from within the company'' through a community of practice (CoP) on the topic of technical debt. This approach can have the advantage that the intrinsic motivation for change is higher and the process is taken more seriously and ``lived'' by the teams using it.
    \end{enumerate} 

    Which path is chosen depends heavily on the corporate culture. However, it is always possible to start with the CoP approach and, if this proves unsuccessful, to try the first approach via management or architects.

\section*{Thanks}
    We would like to thank our industry partners and their team members for their participation, trust, and valuable insights shared with us.
    In memory of Prof. André van Hoorn and Prof. Matthias Riebisch.
    
\section*{Funding and Declaration}
    The project on which this white paper is based was funded by the German Federal Ministry of Education and Research under grant number 01IS24031. 
    The authors are responsible for the content of this publication.
    The opinions expressed herein are my own and do not necessarily represent those of my employer.
    
\section*{Appendix}

    At the end of the document, a collection of screenshots of issue types for technical debt is provided.

\begin{sidewaysfigure*}[!htbp]
            \centering
            \includegraphics[width=0.9\textwidth]{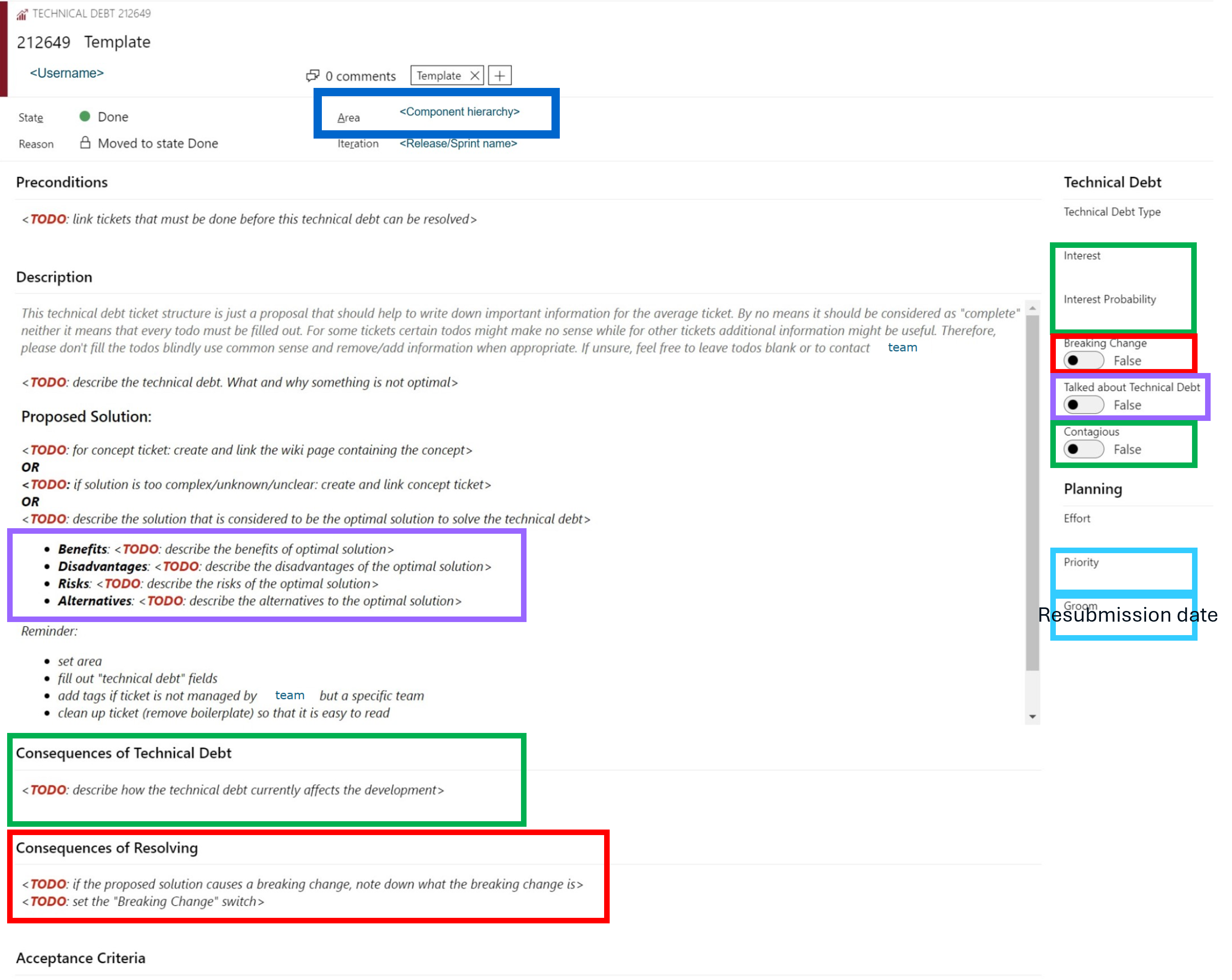}
            \caption{Example of an issue type for technical debt in Azure DevOps. (Purple: Attributes for avoiding technical debt. Green: Attributes for assessing the consequences of technical debt. Red: Attributes for assessing the risks of expanding technical debt. Light blue: Priority and resubmission date for managing technical debt. Dark blue: Component hierarchy for risk assessment of functional changes and dismantling with synergies)}
            \label{fig:Ados}
        \end{sidewaysfigure*}

\begin{sidewaysfigure*}[!htbp]
            \centering
            \includegraphics[width=\textwidth]{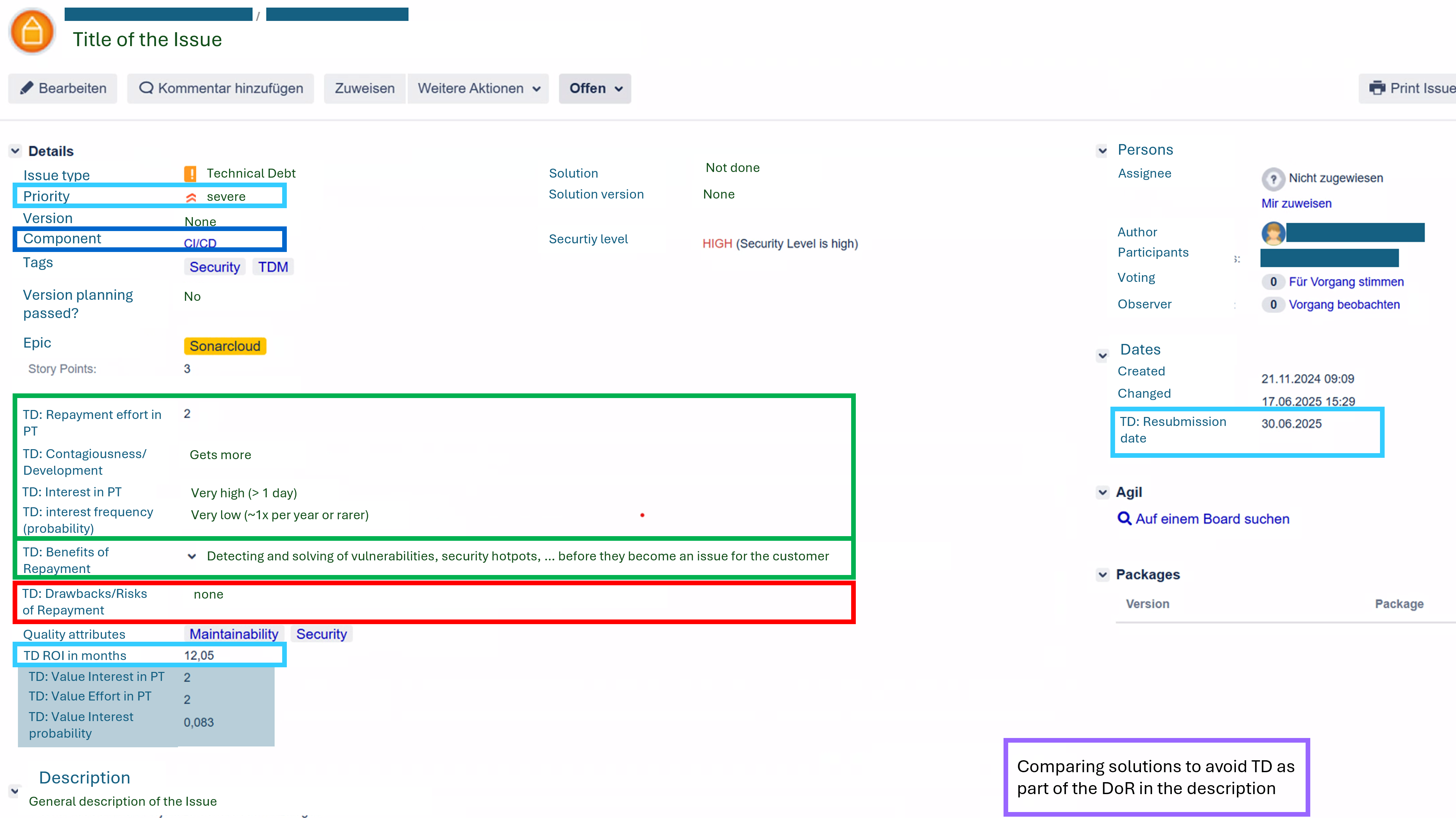}
            \caption{Example of an issue type for technical debt in Jira. (Purple: Attributes for avoiding technical debt. Green: Attributes for assessing the consequences of technical debt. Red: Attributes for assessing the risks of expanding technical debt. Light blue: Priority and resubmission date for managing technical debt. Dark blue: Component hierarchy for risk assessment of functional changes and dismantling with synergies)}
            \label{fig:Jira}
        \end{sidewaysfigure*}

\begin{figure*}[!htbp]
            \centering
            \includegraphics[width=\textwidth]{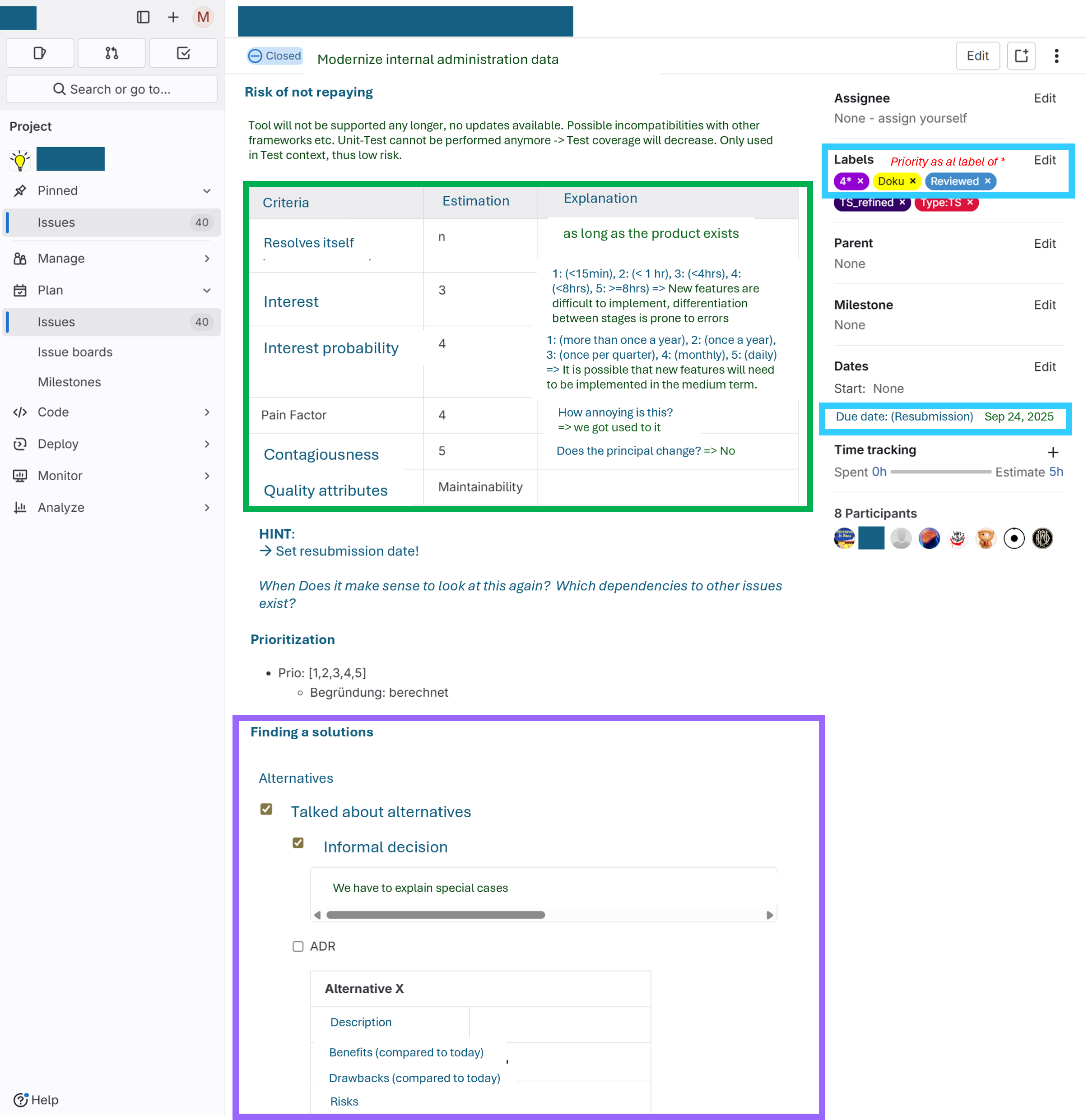}
            \caption{Example of an issue type for technical debt in GitHub (resolved as a template). (Purple: attributes for avoiding technical debt. Green: attributes for assessing the consequences of technical debt. Light blue: priority and resubmission date for managing technical debt. )}
            \label{fig:GitLab}
        \end{figure*}

\begin{enumerate}
    \item Azure DevOps in ~\Cref{fig:Ados}
    \item Jira in ~\Cref{fig:Jira}
    \item GitLab in ~\Cref{fig:GitLab}
\end{enumerate}

\newpage
        
\section*{Acknowledgment}
    We thank the teams that participated in the study for their participation, trust, and insights.
    The project on which this report is based was sponsored by the Federal Ministry of Research, Technology, and Space under the funding code 01IS24031. 
    Responsibility for the content of this publication lies with the authors.

 \bibliography{TDMGuide}

\end{document}